\documentclass[prd,twocolumn,aps, floatfix,preprintnumbers,showpacs,nofootinbib,superscriptaddress]{revtex4-1}
\usepackage{epsfig,natbib}
\usepackage{graphicx}
\usepackage{amsmath}

\newcommand{\be}{\begin{equation}}
\newcommand{\ee}{\end{equation}}
\newcommand{\bea}{\begin{eqnarray}}
\newcommand{\eea}{\end{eqnarray}}
\begin{document}
\title{The Effects of Rayleigh Scattering on the CMB and Cosmic Structure}
\author{Elham Alipour}
\email{elham@phas.ubc.ca}
 \affiliation{Department of Physics and Astronomy, University of British 
Columbia, Vancouver, BC, V6T 1Z1, Canada}
\author{Kris Sigurdson}
 \affiliation{Department of Physics and Astronomy, University of British 
Columbia, Vancouver, BC, V6T 1Z1, Canada}
\author{Christopher M. Hirata}
 \affiliation{Center for Cosmology and Astroparticle Physics, The Ohio State University, 191 West Woodruff Avenue, Columbus, Ohio 43210, USA}
\date{\today}

\begin{abstract}
During and after recombination, in addition to Thomson scattering with free electrons, photons also coupled to neutral hydrogen and helium atoms through Rayleigh scattering. This coupling influences both CMB anisotropies and the distribution of matter in the Universe. The frequency-dependence of the Rayleigh cross section breaks the thermal nature of CMB temperature and polarization anisotropies and effectively doubles the number of variables needed to describe CMB intensity and polarization statistics, while the additional atomic coupling changes the matter distribution and the lensing of the CMB. We introduce a new method to capture the effects of Rayleigh scattering on cosmological power spectra. Rayleigh scattering modifies CMB temperature and polarization anisotropies at the $\sim\!1 \%$ level at $353 {\rm GHz}$ (scaling $\propto \nu^4$), and modifies matter correlations by as much as $\sim\!0.3\%$. We show the Rayleigh signal, especially the cross-spectra between the thermal (Rayleigh) E-polarization and Rayleigh (thermal) intensity signal, may be detectable with future CMB missions even in the presence of foregrounds, and how this new information might help to better constrain the cosmological parameters.
\end{abstract}

\maketitle

\section{Introduction}
Most descriptions of the Cosmic Microwave Background (CMB) anisotropies assume that before recombination at $z\simeq 1090$, photons are tightly coupled to baryons through Thomson scattering with electrons and afterwards  free stream from the surface of last scattering to us \cite{Ma, Mukhanov, Kosowsky}. However, in fact photons were coupled not only to free electrons through Thomson scattering, but also to neutral hydrogen and helium through Rayleigh scattering. The Rayleigh scattering cross section depends approximately on photon frequency to the fourth power and, since it modifies the opacity near decoupling at the few percent level \cite{Peebles}, has been neglected in most of the literature to simplify analysis. In this work we revisit the impact of Rayleigh scattering on cosmological perturbations, quantify its effects, and suggest potential ways that this effect may be detected in the future.

In the past decade the Wilkinson Microwave Anisotropy Probe (WMAP) has provided us with precise measurements of CMB anisotropies \cite{WMAP} and, complemented by next-generation ground based experiments (SPT \cite{SPT}, ACT \cite{ACT}), the Planck satellite has now characterized  the microwave background anisotropies even to a higher precision \cite{PLANCK}. Future measurements may even probe CMB anisotropies with more frequencies and higher precision (e.g., PRISM \cite{PRISM} or PIXIE \cite{PIXIE}). With this dramatic improvement in experimental capability in mind it is timely to include the physics of Rayleigh scattering in  cosmological perturbations theory both to find accurate solutions and forecast whether these effects might be measured with proposed instruments.

A conceptually straightforward method to calculate the effect of Rayleigh scattering on photon perturbations, as its cross section is frequency dependent, is to consider separate Boltzmann hierarchies with different scattering source and visibility functions at each frequency of interest.  While this captures the effects of the extra opacity that photons experience, it does not account for either the momentum transferred to the atoms nor the effect of spectral distortion on gravitational perturbations. In order to model these effects, the photon perturbation at each frequency must be integrated over to determine the photon density and momentum density which influence gravitation perturbations and the photon-baryon coupling. Existing work has modelled the effect of Rayleigh scattering on CMB anisotropies but have avoided determining the baryonic back reaction in detail \cite{Yu, Lewis}. We introduce here a new approach to solve this problem and accurately treat baryons and frequency-dependent photon perturbations simultaneously, allowing us to quantify the impact of Rayleigh scattering on matter perturbations and validate the results of existing CMB anisotropy calculations. The key innovation in our approach is to track perturbations in photon spectral-distortions rather than photon perturbations at a particular frequency.

Rayleigh scattering changes the rate at which photons and baryons decouple from each other, and extra photon drag modifies exactly how baryon perturbations are influenced by photon perturbations. As we quantify below, this alters the shape of the matter correlation function and makes a small shift to the baryon acoustic oscillation (BAO) scale. Like prior work on this subject we find that Rayleigh scattering results in percent level frequency-dependent distortions to CMB power spectra. These distortions break the thermal nature of CMB temperature and polarization anisotropies so that primary CMB intensity and polarization patterns at different frequencies are not perfectly correlated with each other. We show below that to a very good approximation this effectively doubles the number of random variables needed to completely describe the CMB sky, and determine for the first time the set of intensity and E-polarization eigenspectra needed to capture this statistical information. Finally, we forecast how well future CMB missions might detect these eigenspectra and show that a PRISM-like experiment may be able to detect the Rayleigh signal.

The paper is organized as follows: In Section \ref{S_cross_section}, the relevant Rayleigh scattering cross sections for hydrogen and helium are presented. Section \ref{S_Cosmological_equations} reviews the cosmological equations governing the evolution of perturbations in presence of Rayleigh scattering and presents our  new method to calculate the effect of this additional frequency-dependent opacity. The effect of Rayleigh scattering on the matter two-point correlation function and on the CMB power spectra is calculated in Section \ref{S_matter_power_spectrum} and \ref{S_photon_power_spectra} respectively. In Section \ref{S_independent_variables}, we present the two sets of variables needed to describe the CMB intensity and E-polarization statistics. Section \ref{S_detectibility} investigates the possibility of detecting the Rayleigh signal and Section \ref{S_conclusion} concludes.

\section{Rayleigh scattering cross section}
\label{S_cross_section}
The cross section for Rayleigh scattering of a long-wavelength photon from an atom is
\be
\sigma_R(\nu)=\sigma_T|S|^2,
\ee
where $\sigma_T$ is the Thomson cross section and the dimensionless scattering amplitude, S, is given by \cite{Lang}
\be
S=\sum_{j=2} ^\infty f_{1j}\frac{\nu^2}{\nu_{1j}^2-\nu^2}.
\label{CrossSection}
\ee
Here $\nu$ is the photon frequency, $f_{1j}$ is the Lyman series oscillator strength, and $\nu_{1j}$ is the Lyman series frequencies. Note that the summation includes an implied integration over unbound states $j$.

At the time of recombination, when $T\simeq0.25{\rm eV}$, typical photon frequencies are much smaller than $\nu_{1j}$ and it is therefore appropriate to Taylor-expand the dimensionless scattering amplitude as
\be
S=\sum_{k=0}^\infty a_{2k+2}(h\nu)^{2k+2},
\ee
where the coefficients are
\be
a_{2k+2}=\sum_{j\geq 2}f_{1j}(h\nu_{1j})^{-2k-2}+\int_{E_I}^\infty \frac{df}{dE}E^{-2k-2}dE.
\ee
Here we have written the integral over continuum states explicitly. The integral starts at the ionization energy $E_{I}$ of the relevant atom. The Rayleigh scattering cross section is then
\be
\sigma_R=\sigma_T\sum_{k=0}^\infty b_{2k+4}(h\nu)^{2k+4},
\ee 
where 
\be
b_{2k+4}=\sum_{p=0}^k a_{2p+2}a_{2(k-p)+2}.
\ee
The coefficients can be evaluated provided that the oscillator strength distributions are known. For H, these are known exactly: for the discrete spectrum ($1s \rightarrow np$), the oscillator strengths are \cite{Goldwire}
\be
f_{1s,np}=\frac{256n^5(n-1)^{2n-4}}{3(n+1)^{2n+4}},
\ee
with $h\nu_{1s,np}=(1-n^{-2}){\rm Ry}$. Above $E_{I}=1\, {\rm Ry}=13.6{\rm eV}$ there is a continuous spectrum of oscillator strengths,
\be
\frac{df}{dE}=\frac{128e^{-4v{\rm arctan}(v^{-1})}}{3(E/{\rm Ry})^4(1-e^{-2\pi v})}{\rm Ry}^{-1},
\ee
where $v=(E/{\rm Ry}-1)^{-1/2}$ is the principal quantum number of the continuum state.

For He, the electric dipole selection rules allow the ground $1s^2\;^1S$ 
state to have nonzero oscillator strength only with the $^1P$ discrete 
and continuum states. We have taken the oscillator strengths and energies 
for the $1s^2\;^1S\rightarrow 1s\,np\;^1P$ transitions from 
Refs.~\cite{He1, He2} for $n\le 9$ and 
used the asymptotic formula of Ref.~\cite{He3} for $n>9$. For the 
continuum states we used the TOPbase cross sections 
\cite{He4}, which are trivially converted into oscillator 
strengths. The resulting $b_{2k+4}$ coefficients that we adopt for the rest of this work are shown in Table \ref{coefficients}.
\begin{table}
\begin{tabular}{ccccc}
\hline\hline
2k+4& & H & & He  \\
\hline
 4 & &  1.265625 & & 0.120798  \\
 6 & & 3.738281 & & 0.067243  \\
 8 & & 8.813931 & & 0.031585  \\
10 & & 19.153795 & & 0.014153  \\
12 & & 39.923032 & & 0.006226 \\
\hline\hline
\end{tabular}
\caption{The cross-section coefficients $b_{2k+4}{\rm Ry}^{2k+4}$ for H and He in the Rydberg-based units that we adopt for this work.}
\label{coefficients}
\end{table} 
The radiative transfer equations also require the angular distribution and polarization of Rayleigh-scattered radiation. For scattering with initial and final states of zero orbital angular momentum ($S\rightarrow S$), and neglecting spin-orbit coupling, the scattering is of a pure ``scalar'' nature (in the language of Ref.~\cite{1971rqt..book.....B} \S61) and has the same angular and polarization dependence as Thomson scattering, $dP/d\Omega \propto 1+\cos^2\theta$. Near a resonance such as Lyman-$\alpha$, fine structure splitting makes the electron spin important, and the scattering by hydrogen takes on a different form that is a combination of scalar, anti-symmetric, and symmetric scattering; the full equations for the angular scattering distribution as a function of frequency through the resonance can be found in e.g. Appendix B of Ref.~\cite{2006MNRAS.367..259H}. The equations in Appendix B of Ref.~\cite{2006MNRAS.367..259H} show that the angular distribution approaches the scalar case with corrections of order $\Delta\nu_{\rm fs}^2/(\nu_{{\rm Ly}\alpha}-\nu)^2$ as one moves away from the resonance, where the fine structure splitting is $\Delta\nu_{\rm fs}\sim 11\,$GHz. For cases considered in this paper (frequencies up to 857 GHz observer frame, or $0.52\nu_{{\rm Ly}\alpha}$ at $z=1500$), we are thus safely below the lowest resonant frequency, and the scalar angular distribution -- already incorporated in the CMB Boltzmann hierarchy formalism -- is applicable.

\section{Cosmological equations}
\label{S_Cosmological_equations}
\begin{figure}
\includegraphics[scale=0.7]{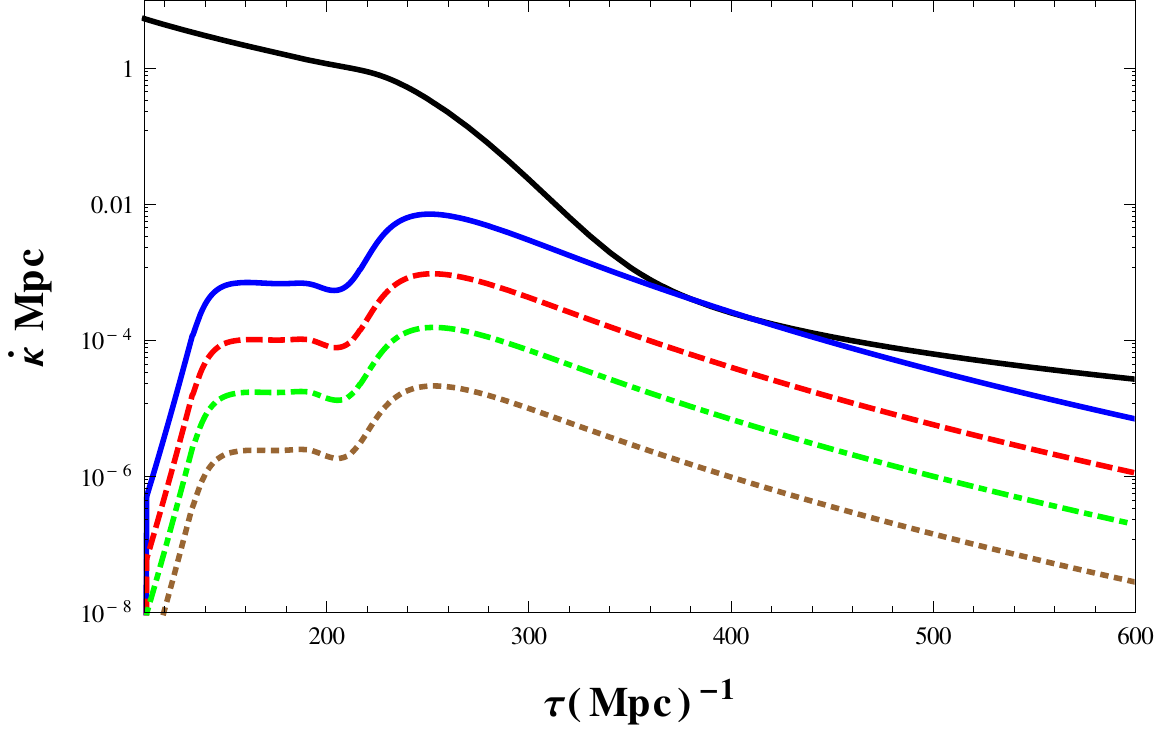}
\centering
\caption{The Comoving opacity as a function of comoving time. The black line is for Thomson scattering while the blue, red, green and brown lines are for Rayleigh scattering at frequencies 857, 545, 353, and 217 GHz respectively.}
\label{opacity}
\end{figure}

To include the effects of Rayleigh scattering on cosmological perturbations, we must modify the evolution equations for photon temperature, photon polarization  and baryon velocity perturbations. We use synchronous gauge in this paper as it is convenient for most numerical computations. The full cosmological evolution equations in this gauge are given in a number of papers \cite{Ma, Challinor}, and therefore we  only explicitly show the equations that need modification. In particular, using the Boltzmann equation in this gauge we find the evolution equations for the temperature perturbation, $\Theta_I$, and E-polarization, $\Theta_E$, hierarchies are
\bea
\dot{\Theta}_{I0}&=&-k\Theta_{I1}+\frac{\dot{a}}{a}\nu\frac{\partial{\Theta_{I0}}}{\partial{\nu}}-\frac{\dot{h}}{6},\\
\dot{\Theta}_{I1}&=&\frac{k}{3}\Theta_{I0}-\frac{2k}{3}\Theta_{I2}+\frac{\dot{a}}{a}\nu\frac{\partial{\Theta_{I1}}}{\partial{\nu}}\nonumber\\
&&-\dot{\kappa}[-\Theta_{I1}+\frac{1}{3}v_b],\\
\dot{\Theta}_{I2}&=&\frac{2k}{5}\Theta_{I1}-\frac{3k}{5}\Theta_{I3}+\frac{\dot{a}}{a}\nu\frac{\partial{\Theta_{I2}}}{\partial{\nu}}+\frac{\dot{h}+6\dot{\eta}}{15}\nonumber\\
&&-\dot{\kappa}[-\Theta_{I2}+\frac{1}{10}\Pi],\\
\dot{\Theta}_{Il}&=&\frac{k}{2l+1}[l\Theta_{I(l-1)}-(l+1)\Theta_{I(l+1)}]\nonumber\\
&&+\frac{\dot{a}}{a}\nu\frac{\partial{\Theta_{Il}}}{\partial{\nu}}+\dot{\kappa}\Theta_{Il}\hspace{2.2cm} l\geq 3, \\
\dot{\Theta}_{E2}&=&\frac{2k}{5}\Theta_{E1}-\frac{k}{3}\Theta_{E3}+\frac{\dot{a}}{a}\nu\frac{\partial{\Theta_{E2}}}{\partial{\nu}}\nonumber\\
&&+\dot{\kappa}(\Theta_{E2}-\frac{2}{5}\Pi),\\
\dot{\Theta}_{El}&=&\frac{k}{2l+1}[l\Theta_{E(l-1)}-\frac{(l+3)(l-1)}{l+1}\Theta_{E(l+1)}]\nonumber\\
&&+\frac{\dot{a}}{a}\nu\frac{\partial{\Theta_{El}}}{\partial{\nu}}+\dot{\kappa}\Theta_{El} \hspace{2cm} l\geq 3,
\eea
where an overdot denotes derivatives with respect to conformal time $\tau$, k is the wavenumber of the perturbations, $h$ and $\eta$ are the synchronous gauge metric perturbations, $\Pi$ is the combination 
$\Theta_{I2}+\frac{3}{2}\Theta_{E2}$, $a$  the scale factor and $\dot{\kappa}$ is the comoving opacity defined as
\bea
-\dot{\kappa}&=&-\dot{\kappa}_T-\dot{\kappa}_R\nonumber\\
&=&n_e\sigma_Ta+n_H\sigma_R^Ha+n_{He}\sigma_R^{He}a.
\eea
Here $n_e$, $n_H$ and $n_{He}$ are respectively the number densities of free electrons, neutral hydrogen and helium atoms. The comoving opacity for Rayleigh and Thomson scattering as a function of conformal time is plotted in Figure \ref{opacity} for a couple of observed frequencies. 

In standard case when opacity does not depend on frequency, the baryons evolve according to equations 
\be
\dot{\delta_b}=-kv_b-\frac{1}{2}\dot{h},
\ee
\bea
\dot{v_b}+\frac{\dot{a}}{a}v_b-kc_s^2\delta_b &=& \frac{1}{\bar{\rho}_b}\int \frac{d^3p}{(2\pi)^3}(-p\mu)C[f(\vec{p})]\nonumber\\
&=& \frac{4\bar{\rho}_\gamma}{3\bar{\rho}_b}\dot{\kappa}(-3\Theta_{I1}+v_b),
\eea
where $\delta_b$ and $v_b$ are baryon overdensity and velocity, $c_s$ is the intrinsic baryon sound speed, $f(p)$ is photon distribution function, $C[f(\vec{p})]=\frac{df}{dt}$ is the collision term in the Boltzmann equation for photon temperature perturbations, $\mu=\hat{p}\cdot\hat{k}$, and $\bar{\rho}_\gamma$ and $\bar{\rho}_b$ are the mean photon and baryon energy densities.

Including Rayleigh scattering will make the opacity frequency-dependent, therefore the scattering term in the baryon velocity must be modified to
\bea
&&\int \frac{d^3p}{(2\pi)^3}(-p\mu)C[f(\vec{p})]=\nonumber \\
&&\int \frac{d^3p}{(2\pi)^3}p^2\frac{\partial f}{\partial p} \mu \dot{\kappa}(p)(\Theta_{I0}(p)-\Theta_{I}(p)+\mu v_b).
\eea

As discussed above, a straightforward method to solve the above system of equations is to consider a separate Boltzmann hierarchy for each frequency of interest, each with different scattering sources and visibility function, and then integrate over each photon frequency bin to get the total baryon-photon coupling \cite{Yu, Lewis}. However there is another computationally efficient method that can be used. If at the times that atoms are present the typical CMB photon energies are much smaller than Rydberg energy $h\nu\ll {\rm Ry}$, then we can write $\Theta_{Il}$ and $\Theta_{El}$ as Taylor series in the comoving frequency $a\nu$ where each term in the series describes spectral-distortion perturbations that scale with increasing powers of frequency. Specifically we write:
\be
\Theta_{Il}(\nu)=\sum_{r=0}^\infty \Theta_{Il}^{(2r)}\left(\frac{ah\nu}{a_{*}{\rm Ry}}\right)^{2r},
\label{Taylor1}
\ee
\be
\Theta_{El}(\nu)=\sum_{r=0}^\infty \Theta_{El}^{(2r)}\left(\frac{ah\nu}{a_{*}{\rm Ry}}\right)^{2r}.
\label{Taylor2}
\ee
Note that only even powers of $\nu$  appear because the scattering cross section contains only even powers of $\nu$. We expanded the perturbations in terms of $ah\nu/a_{*}{\rm Ry}$ because this ratio does not evolve with time for a given photon and $a_{*}=0.001$ is a reference epoch for normalizing the coefficients in the series expansion (its value has no physical consequences). Similarly we can write the opacity as
\be
\dot{\kappa}(\nu)=\sum_{r=0}^\infty \dot{\kappa}_{2r}\left(\frac{ah\nu}{a_{*}{\rm Ry}}\right)^{2r},
\ee
where $\dot{\kappa}_0=-n_e\sigma_Ta$ is the standard Thomson scattering rate, $\dot{\kappa}_1=0$ and 
\be
-\dot{\kappa}_{2r}=(n_Hb_{2r}^H+n_{He}b_{2r}^{He})\sigma_Ta\left(\frac{a_{*}}{a}\right)^{2r}.
\ee

Substituting  these Taylor expansions into evolution equations for photon temperature and polarization perturbations leads to the following evolution equations for each $\Theta_{Il}^{(2n)}$ and $\Theta_{El}^{(2n)}$ terms:
\bea
\dot{\Theta}_{I0}^{(2n)}&=&-k\Theta_{I1}^{(2n)}-\frac{\dot{h}}{6}\delta_{n,0},\\
\dot{\Theta}_{I1}^{(2n)}&=&\frac{k}{3}\Theta_{I0}^{(2n)}-\frac{2k}{3}\Theta_{I2}^{(2n)}\nonumber\\
&&-\sum_{r=0}^n \dot{\kappa}_{2r} [-\Theta_{I1}^{2(n-r)}+\frac{v_b}{3}\delta_{n-r,0}],\\
\dot{\Theta}_{I2}^{(2n)}&=&\frac{2k}{5}\Theta_{I1}^{(2n)}-\frac{3k}{5}\Theta_{I3}^{(2n)}+\frac{\dot{h}+6\dot{\eta}}{15}\delta_{n,0}\nonumber\\
&&-\sum_{r=0}^n \dot{\kappa}_{2r} [-\Theta_{I2}^{2(n-r)}+\frac{\Pi^{2(n-r)}}{10}],\\
\dot{\Theta}_{Il}^{(2n)}&=&\frac{k}{2l+1}[l\Theta_{I(l-1)}^{(2n)}-(l+1)\Theta_{I(l+1)}^{(2n)}]\nonumber\\
&&+\sum_{r=0}^n \dot{\kappa}_{2r} \Theta_{Il}^{2(n-r)}\hspace{2cm} l\geq 3,
\eea
\bea
\dot{\Theta}_{E2}^{(2n)}&=&\frac{2k}{5}\Theta_{E1}^{(2n)}-\frac{k}{3}\Theta_{E3}^{(2n)}\nonumber\\
&&+\sum_{r=0}^n \dot{\kappa}_{2r}[\Theta_{E2}^{2(n-r)}-\frac{2}{5}\Pi^{2(n-r)}],\\
\dot{\Theta}_{El}^{(2n)}&=&\frac{k}{2l+1}[l\Theta_{E(l-1)}^{(2n)}-\frac{(l+3)(l-1)}{l+1}\Theta_{E(l+1)}^{(2n)}]\nonumber\\
&&+\sum_{r=0}^n \dot{\kappa}_{2r}\Theta_{El}^{2(n-r)}\hspace{2cm} l\geq 3.
\eea
To find the evolution equation for baryon velocity we first must calculate the following integral
\bea
I_n&=&-\frac{1}{4\bar{\rho}_\gamma T^n}\int_0^\infty \frac{d\nu}{2\pi^2} \nu^{n+4}\frac{\partial f}{\partial \nu}\nonumber \\
&=&\frac{15}{4\pi^4}(n+4)!\zeta[n+4],
\eea
where $\zeta$ is the Riemann $\zeta$-function. Therefor the baryon velocity in the presence of Rayleigh scattering evolves according to
\bea
\dot{v}_b&=&-\frac{\dot{a}}{a}v_b+kc_s^2\delta_b\nonumber\\
&&+\frac{4\bar{\rho}_\gamma}{3\bar{\rho}_b}\sum_{r=0}^\infty \dot{\kappa}_{2r}[-3\sum_{n=0}^\infty\Theta_{I1}^{(2n)}I_{2(n+r)}\left(\frac{aT}{a_{*}{\rm Ry}}\right)^{2(n+r)}\nonumber\\
&&+v_bI_{2r}\left(\frac{aT}{a_{*}{\rm Ry}}\right)^{2r}].
\eea

As shown in Equation \ref{CrossSection}, the Rayleigh cross section blows up near the resonant frequencies. Therefore photons with these frequencies remain tightly coupled to baryons. Photons do not self interact so these resonant photons are unlikely to change the CMB power spectrum. However they do enhance the pressure or sound speed of baryons. There is typically of order 1 photon per baryon near the Lyman-$\alpha$ line and since the photon energy is 10.2 eV, and the baryon mass is 1 GeV, the baryon sound speed increases by roughly $10^{-8}$. This only alters perturbations at very small scales below those of interest in this work. 

Since metric perturbation evolution depends on the total photon overdensity and velocity, the final modification is to calculate the change in the photon stress-energy tensor in the presence of frequency dependent scattering. The fractional photon energy density perturbation is
\bea
\delta_\gamma&=& -\frac{1}{\bar{\rho}_\gamma}\int{\nu^4}d\nu\frac{\partial f}{\partial \nu}\Theta_{I0}(\nu)\nonumber\\
&=&4\sum_{r=0}^\infty\Theta_{I0}^{(2r)}I_{2r}\left(\frac{aT}{a_{*}{\rm Ry}}\right)^{2r},
\eea
and the photon momentum density is
\bea
\Theta_\gamma&=&-\frac{3k}{4\bar{\rho}_\gamma}\int{\nu^4}d\nu\frac{\partial f}{\partial \nu}\Theta_{I1}(\nu)\nonumber\\
&=&3k\sum_{r=0}^\infty\Theta_{I1}^{(2r)}I_{2r}\left(\frac{aT}{a_{*}{\rm Ry}}\right)^{2r}.
\eea

This appears to replace the problem of summing over many perturbations at different frequencies with summing over many  perturbations with different  spectral-distortion shapes.
However, we find in practice that these sums rapidly converge after including only a few of the spectral-distortion terms which allows the entire system to be solved for efficiently and accurately.

\section{Matter power spectrum}
\label{S_matter_power_spectrum}
One of the physical effects of Rayleigh scattering is a change in matter two-point correlation function. The matter correlation function is the excess probability, compared with what expected from a random distribution, of finding a matter over-density at a distance $\vec{r}$ apart and its Fourier transform is the matter power spectrum,
\be
\xi(\vec{r})=\langle\delta(\vec{x})\delta(\vec{x}+\vec{r})\rangle=\int \frac{d^3k}{(2\pi)^3}P(k)e^{i\vec{k}.\vec{r}}.
\ee

\begin{figure}
\includegraphics[width=0.43\textwidth]{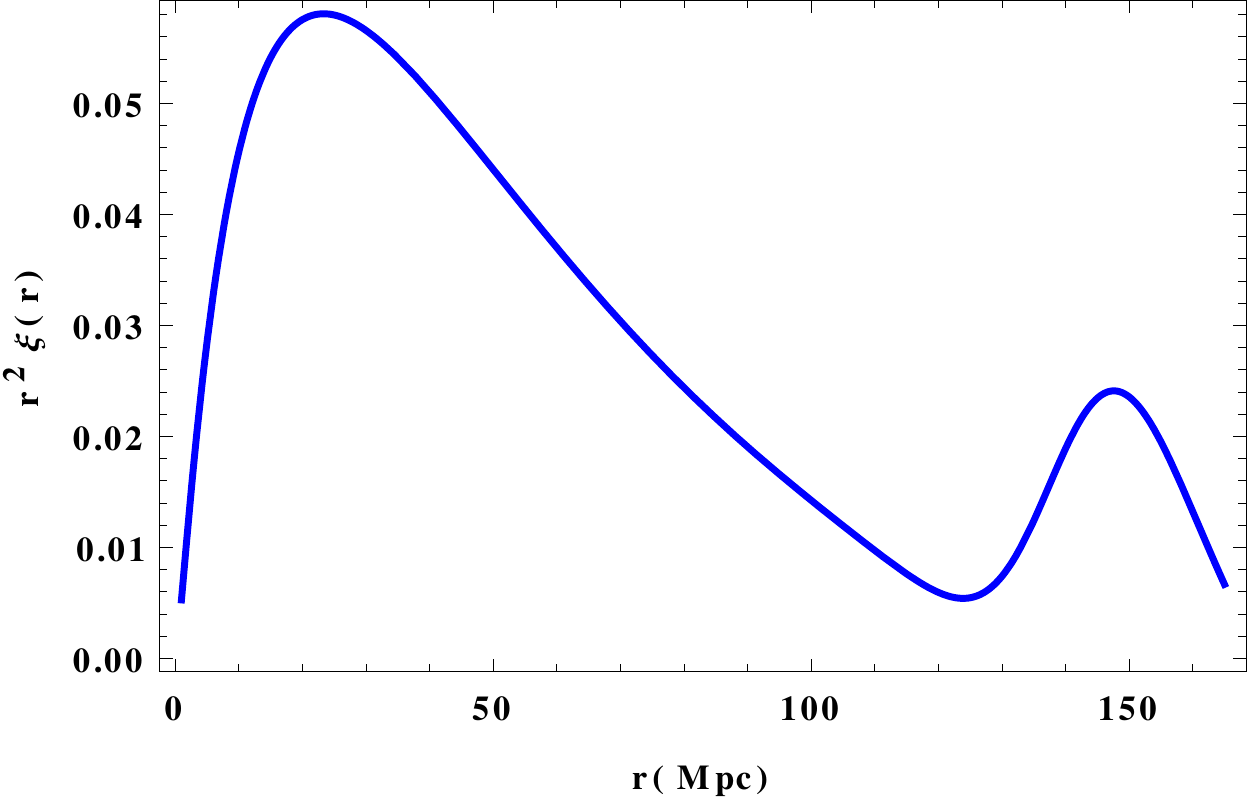}
\centering
\caption{The matter two-point correlation function, $r^2\xi(\vec{r})$, as a function of the distance between two over-densities for our fiducial cosmological parameters.}
 \label{Corr}
\end{figure}

\begin{figure}
\includegraphics[width=0.43\textwidth]{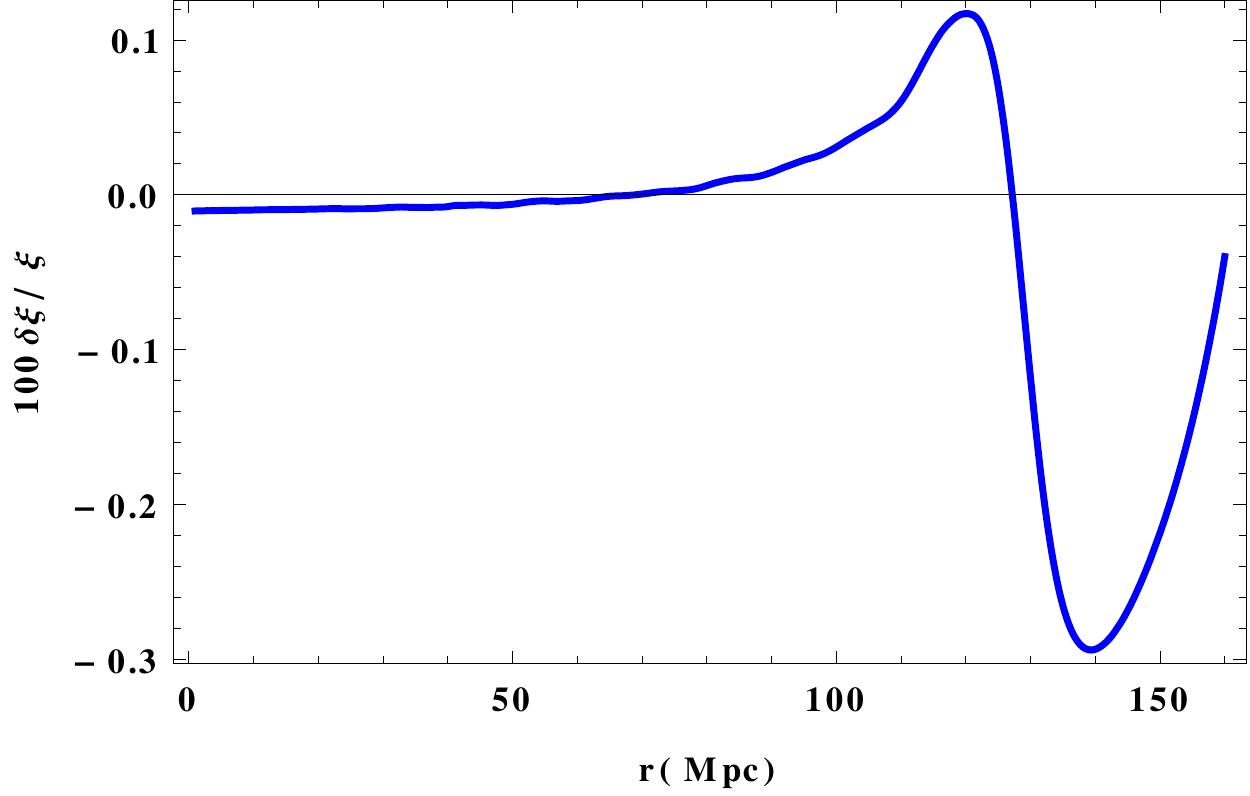}
\centering
\caption{The percentage change in the matter correlation function due to Rayleigh scattering for our fiducial cosmological parameters.}
 \label{DiffCorr}
\end{figure}

Rayleigh scattering increases the total baryon-photon coupling which delays the time of recombination. As shown in Figure \ref{Corr}, the correlation function has a peak near a radius of $\sim150$\,{\rm Mpc}, the BAO scale, which represent the sound horizon at the time of recombination. This changes due to the delay in the time of photon-baryon decoupling. The percentage change in the two-point correlation function due to Rayleigh scattering is plotted in Figure \ref{DiffCorr}. Adding Rayleigh scattering to the opacity changes the correlation function by up to $\sim0.3\%$. Unless otherwise stated we show all results in a fiducial model where we adopt the best-fit parameters from PLANCK \cite{PLANCK}.

\begin{figure*}
\includegraphics[width=\textwidth, height=0.645\textwidth]{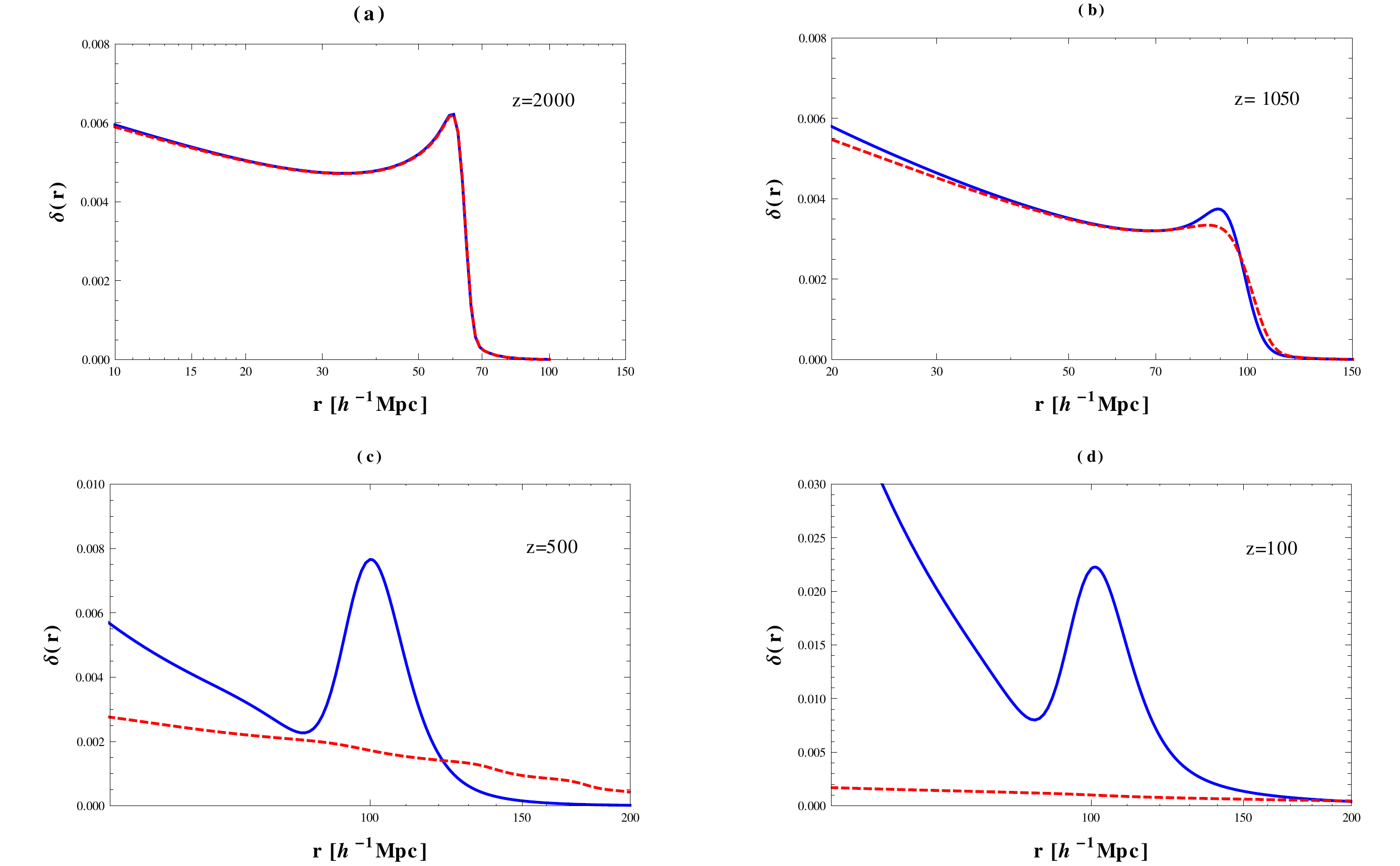}
\centering
\caption{The redshift evaluation of a narrow Gaussian-shaped adiabatic density fluctuation in real space. The blue and red lines are respectively the baryon and photon density waves. At very early times, (panel (a)), baryons and photons are tightly coupled and their density waves travel together. As time goes by, (panels (b) to (d)), they decouple, photons free stream to us and baryons cluster around the initial over-density and in a shell at about 150 Mpc radius.}
\label{real-space}
\end{figure*}

Another way of visualizing how much the matter power spectrum is changed in the presence of Rayleigh scattering is by looking at the evolution of a concentrated matter over-density in real space. In Figure \ref{real-space}, the redshift evaluation of a narrow Gaussian-shaped adiabatic density fluctuation in real space is displayed. 

At very early stages, when the photons and baryons were tightly coupled, panel (a), the baryon-photon plasma density wave travels outward from the initial over-density. Panel (b) shows a snapshot of the density waves at redshift $z=1050$. At this time the temperature is low enough that the neutral atoms can form, therefore the photons begin to decouple from baryons and the sound speed starts to drop. Thus the baryon density wave slows down compare to photon density wave. In panel (c), the waves are shown at $z=500$ when photons and baryons are completely decoupled. The photon perturbation smooths itself out at the speed of light. But because the sound speed is much smaller than speed of light the baryon density wave stalls. Panel (d) present the late time picture. The photons free-stream until now when we can observe them as the cosmic microwave background and the baryon perturbation clusters around the initial over-density and in a shell about $\sim 150$ Mpc radius.

\begin{figure}
\includegraphics[width=0.45\textwidth, height=0.3\textwidth]{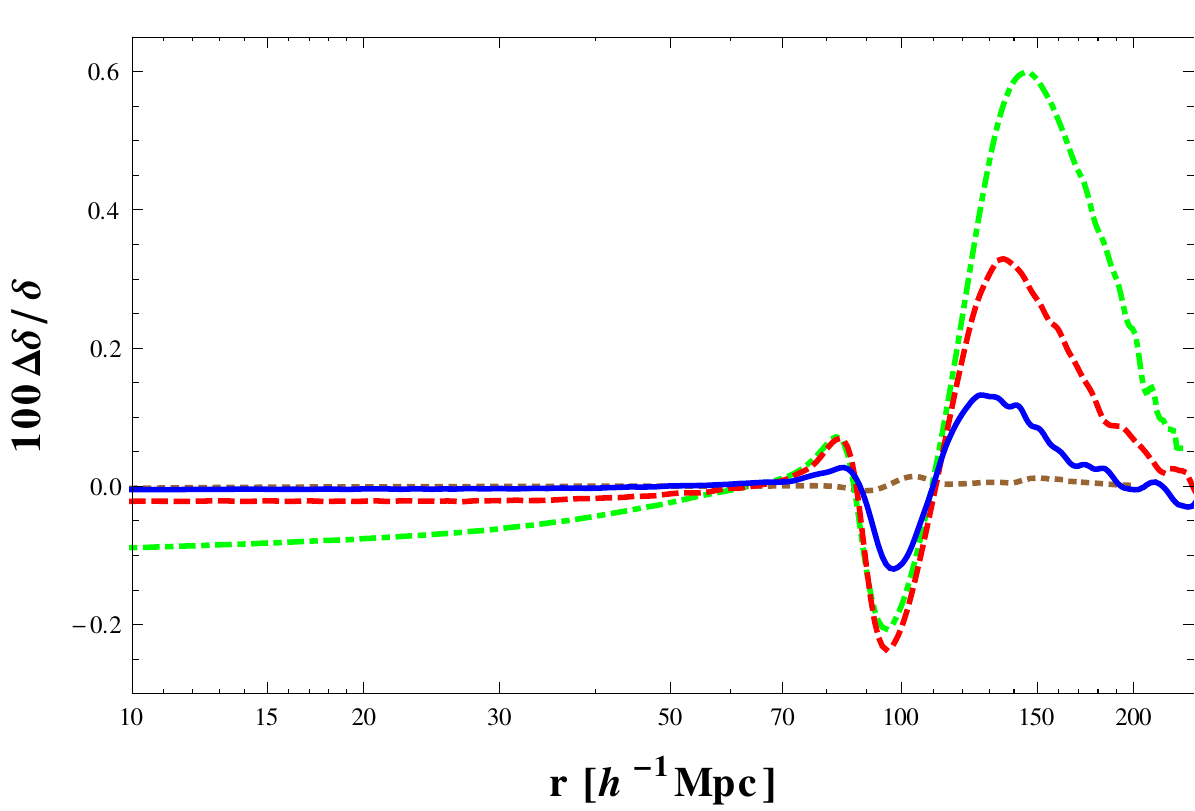}
\centering
\caption{The percentage change in physical baryon density fluctuations in real space due to Rayleigh scattering at different redshifts. The blue, red, green and brown lines correspond to redshifts 0, 100, 500 and 1050 respectively.}
\label{diff-real-space}
\end{figure}

In Figure \ref{diff-real-space}, the percentage change in physical baryon density fluctuations in real space due to Rayleigh scattering is plotted at different redshifts. Note that while $\Delta\delta/\delta$ is up to $0.6\%$ at some points the percentage change in the location of the peak in baryon density wave or the BAO scale due to Rayleigh scattering is less than 
$0.01\%$ in this example, and so the detailed effect of Rayleigh scattering is not well modelled as a simple shift in the BAO scale.

\section{Photon power spectra}
\label{S_photon_power_spectra}
\begin{figure}
\includegraphics[width=0.45\textwidth, height=0.28\textwidth]{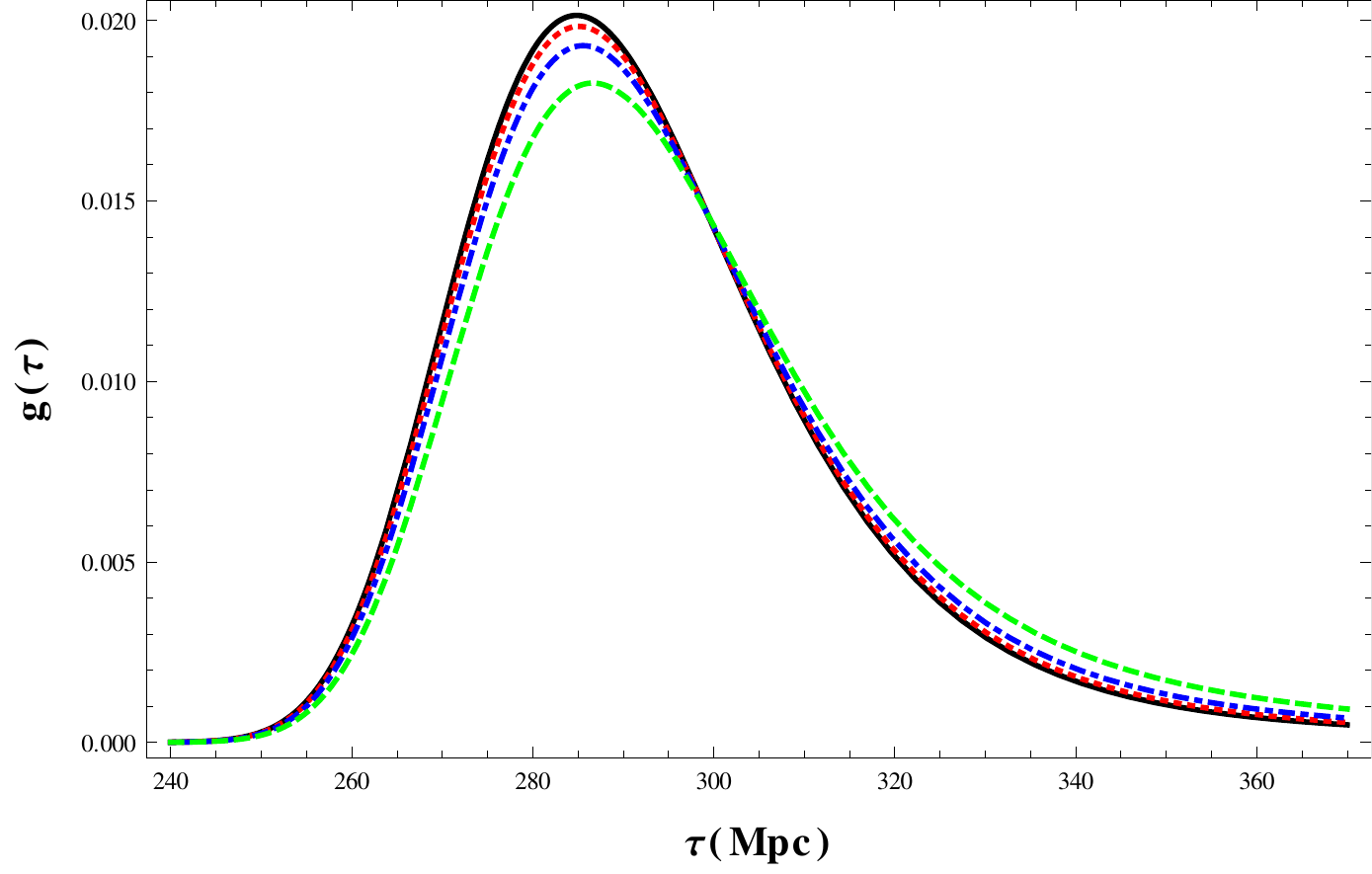}
\centering
\caption{The total visibility function as a function of conformal time for several frequencies. The black, red, blue and green lines are the total visibility function for frequencies 0, 545, 700 and 857 GHz respectively. The total photon visibility function shifts toward later times with increasing frequency.}
\label{vis}
\end{figure}

To calculate the power spectra for both photon temperature and E-polarization perturbations, we use the line of sight integration approach of Ref.~\cite{Seljak}. In this approach, the solutions of Eqs. (23-28) can be written as an integral over the product of a source term and a geometrical term which is just the spherical Bessel function,
\bea
\Theta_{Il}(\tau_0)&=&\int_0^{\tau_0}d\tau S_I(k,\tau)j_l[k(\tau_0-\tau)],\\
\Theta_{El}(\tau_0)&=&\int_0^{\tau_0}d\tau S_E(k,\tau)j_l[k(\tau_0-\tau)].
\eea 
The source functions for temperature and E-polarization perturbations are given in many previous studies \cite{Seljak, Challinor}. 
\bea
S_{I}(k,\tau)&=&e^{-\kappa}[-\frac{\dot{h}}{6}+\frac{k}{3}\sigma+\frac{\ddot{\sigma}}{k}]\\
&&+g(\tau)[2\frac{\dot{\sigma}}{k}+\Theta_{I0}+\frac{\dot{v_b}}{k}+\frac{\Pi}{4}+\frac{3}{4k^2}\ddot{\Pi}]\nonumber\\
&&+\dot{g}(\tau)[\frac{\sigma}{k}+\frac{v_b}{k}+\frac{3}{4k^2}2\dot{\Pi}]+\ddot{g}(\tau)\frac{3}{4k^2}\Pi^{(0)},\nonumber\\
S_E(k,\tau)&=&g(\tau)\frac{3}{4}\Pi\frac{1}{[k(\tau_0-\tau)]^2},
\eea
where $\sigma=(\dot{h}+6\dot{\eta})/2k$ and $g(\tau)=-\dot{\kappa}e^{-\kappa}$ is the visibility function. In the presence of Rayleigh scattering the visibility function is frequency dependent and can be written as a Taylor series in $ah\nu/a_{*}{\rm Ry}$. The total visibility function for several frequencies is plotted in Figure \ref{vis}. Note that the total photon visibility function shifts toward later time with increasing frequencies.

Substituting the Taylor expansions of visibility function and temperature and E-polarization perturbations into the above equations gives the source functions for each of the  $\Theta_{Il}^{(2n)}$ and $\Theta_{El}^{(2n)}$ terms,
\bea
\Theta_{Il}^{(2n)}(\tau_0)&=&\int_0^{\tau_0}d\tau S_I^{(2n)}(k,\tau)j_l[k(\tau_0-\tau)],\\
\Theta_{El}^{(2n)}(\tau_0)&=&\int_0^{\tau_0}d\tau S_E^{(2n)}(k,\tau)j_l[k(\tau_0-\tau)],
\eea 
where
\begin{widetext}
\bea
S_I^{(0)}&=&e^{-\kappa_0}[-\frac{\dot{h}}{6}+\frac{k}{3}\sigma+\frac{\ddot{\sigma}}{k}]+g_0[2\frac{\dot{\sigma}}{k}+\Theta_{I0}^{(0)}+\frac{\dot{v_b}}{k}+\frac{\Pi^{(0)}}{4}+\frac{3}{4k^2}\ddot{\Pi}^{(0)}]\nonumber\\
&&+\dot{g}_0[\frac{\sigma}{k}+\frac{v_b}{k}+\frac{3}{4k^2}2\dot{\Pi}^{(0)}]+\ddot{g}_0\frac{3}{4k^2}\Pi^{(0)},\\
S_I^{(4)}&=&e^{-\kappa_0}[-\frac{\dot{h}}{6}+\frac{k}{3}\sigma+\frac{\ddot{\sigma}}{k}](-\kappa_4)+(g_0(-\kappa_4)+g_4)[2\frac{\dot{\sigma}}{k}+\Theta_{I0}^{(0)}+\frac{\dot{v_b}}{k}+\frac{\Pi^{(0)}}{4}+\frac{3}{4k^2}\ddot{\Pi}^{(0)}]\nonumber\\
&&+g_0[\Theta_{I0}^{(4)}+\frac{\Pi^{(4)}}{4}+\frac{3}{4k^2}\ddot{\Pi}^{(4)}]+(\dot{g}_0(-\kappa_4)+g_0(-\dot{\kappa}_4)+\dot{g}_4)[\frac{\sigma}{k}+\frac{v_b}{k}+\frac{3}{4k^2}2\dot{\Pi}^{(0)}]\nonumber\\
&&+\dot{g}_0\frac{3}{4k^2}2\dot{\Pi}^{(4)}+\ddot{g}_0\frac{3}{4k^2}\Pi^{(4)}+[\ddot{g}_0(-\kappa_4)+2\dot{g}_0(-\dot{\kappa}_4) +g_0(-\ddot{\kappa}_4)+\ddot{g}_4]\frac{3}{4k^2}2\Pi^{(0)},\\
S_I^{(6)}&=&e^{-\kappa_0}[-\frac{\dot{h}}{6}+\frac{k}{3}\sigma+\frac{\ddot{\sigma}}{k}](-\kappa_6)+(g_0(-\kappa_6)+g_6)[2\frac{\dot{\sigma}}{k}+\Theta_{I0}^{(0)}+\frac{\dot{v_b}}{k}+\frac{\Pi^{(0)}}{4}+\frac{3}{4k^2}\ddot{\Pi}^{(0)}]\nonumber\\
&&+g_0[\Theta_{I0}^{(6)}+\frac{\Pi^{(6)}}{4}+\frac{3}{4k^2}\ddot{\Pi}^{(6)}]+(\dot{g}_0(-\kappa_6)+g_0(-\dot{\kappa}_6)+\dot{g}_6)[\frac{\sigma}{k}+\frac{v_b}{k}+\frac{3}{4k^2}2\dot{\Pi}^{(0)}]\nonumber\\
&&+\dot{g}_0\frac{3}{4k^2}2\dot{\Pi}^{(6)}+\ddot{g}_0\frac{3}{4k^2}\Pi^{(6)}+[\ddot{g}_0(-\kappa_6)+2\dot{g}_0(-\dot{\kappa}_6) +g_0(-\ddot{\kappa}_6)+\ddot{g}_6]\frac{3}{4k^2}2\Pi^{(0)},\\
S_E^{(0)}&=&\frac{3}{4[k(\tau_0-\tau)]^2}g_0\Pi^{(0)},\\
S_E^{(4)}&=&\frac{3}{4[k(\tau_0-\tau)]^2}(g_0[\Pi^{(4)}+\Pi^{(0)}(-\kappa_4)]+g_4\pi^{(0)}),\\
S_E^{(6)}&=&\frac{3}{4[k(\tau_0-\tau)]^2}(g_0[\Pi^{(6)}+\Pi^{(0)}(-\kappa_6)]+g_6\pi^{(0)}).
\eea
\end{widetext}

Here $g_{2r}=-\dot{\kappa}_{2r}e^{-\kappa_0}$. The anisotropy spectrum can be obtained by integrating over the initial power spectrum of the metric perturbation, $P_\psi(k)$:
\bea
C_l^{XY}(\nu,\nu')&=&\int_0^\infty k^2dkP_\psi(k)\left(\Theta_{Xl}(\nu,k)\Theta_{Yl}(\nu',k)\right)\\
&=&\sum_{r,r'=0}^\infty C_l^{XY(2r,2r')}\left(\frac{ah\nu}{a_{*}{\rm Ry}}\right)^{2r}\left(\frac{ah\nu'}{a_{*}{\rm Ry}}\right)^{2r'},\nonumber
\label{clXY}
\eea
where 
\be
C_l^{XY(2r,2r')}=\int_0^\infty k^2dkP_\psi(k)(\Theta_{Xl}^{(2r)}(k)\Theta_{Yl}^{(2r')}(k)).
\ee
\begin{figure*}
\includegraphics[width=\textwidth, height=0.4\textwidth]{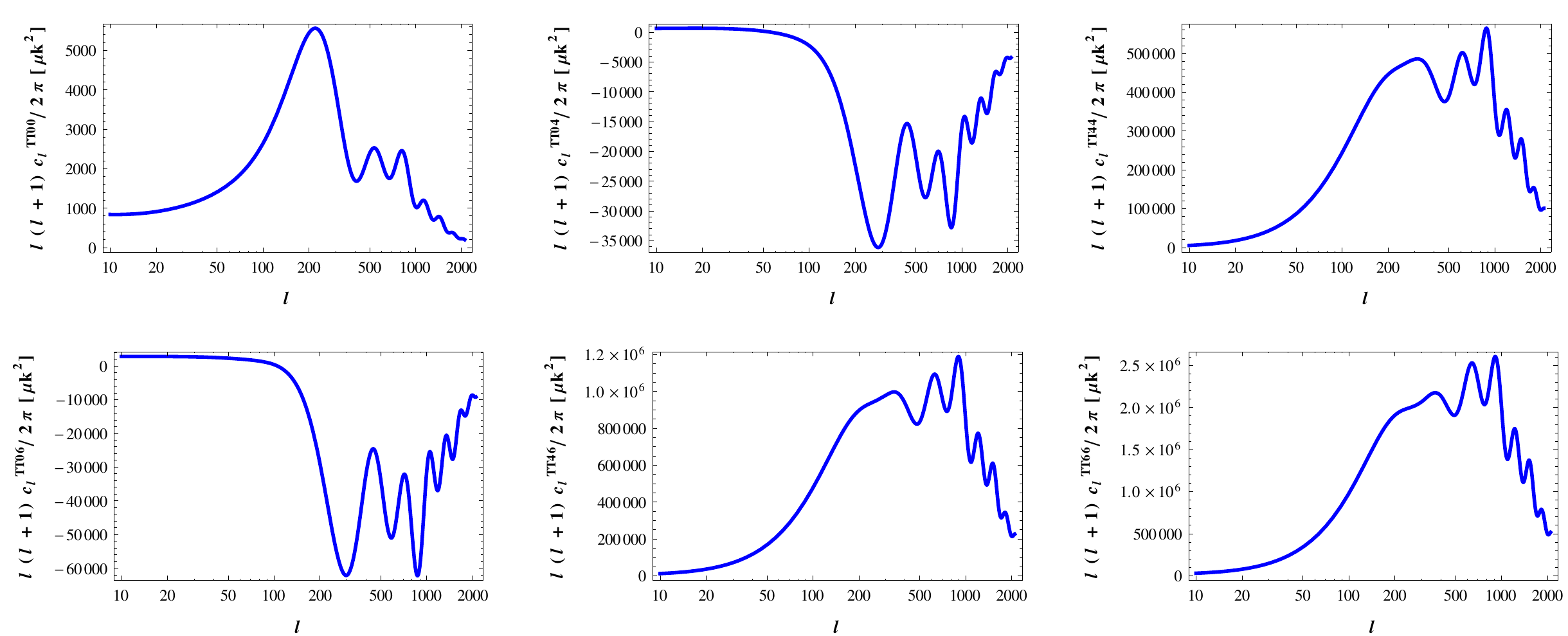}
\centering
\caption{The cross correlation temperature power spectrum $C_l^{TT(2r,2r')}$ of the $\Theta_{Il}^{(2r)}$ and $\Theta_{Il}^{(2r')}$ intensity coefficients for the $\nu^0$, $\nu^4$ and $\nu^6$ spectral distortions.}
\label{clTTnm}
\end{figure*}

\begin{figure*}
\includegraphics[width=\textwidth, height=0.4\textwidth]{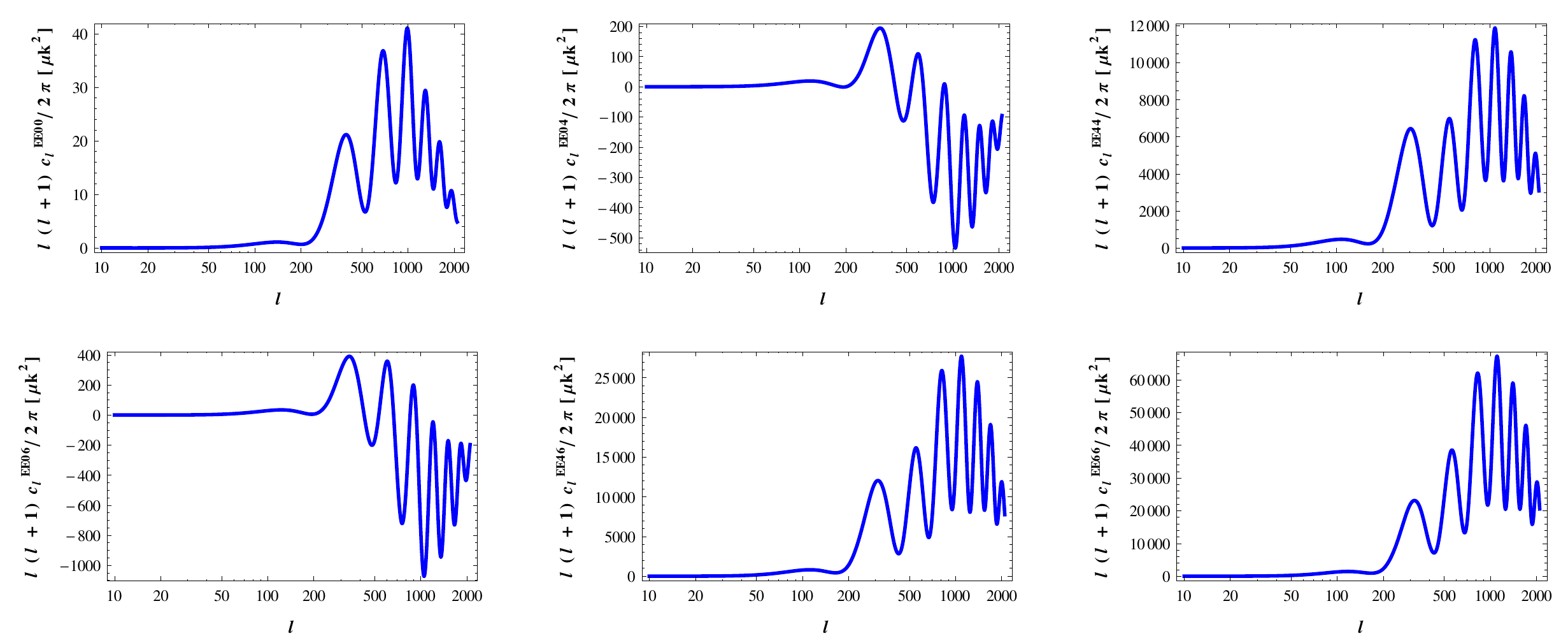}
\centering
\caption{The cross correlation temperature power spectrum $C_l^{EE(2r,2r')}$ of the $\Theta_{El}^{(2r)}$ and $\Theta_{El}^{(2r')}$ E-polarization coefficients for the $\nu^0$, $\nu^4$ and $\nu^6$ spectral distortions.}
\label{clEEnm}
\end{figure*}
We used a modified version of CAMB \cite{CAMB} to numerically calculate $C_l^{TT(2r,2r')}$ and $C_l^{EE(2r,2r')}$ power spectra. These results are shown in Figure \ref{clTTnm} and Figure \ref{clEEnm}. Note that while Eq. \ref{clXY} describes unlensed power spectra from the surface of last scattering, here and elsewhere, these power spectra include the effect of gravitational lensing from structure along the line of sight implemented in CAMB. 

\begin{figure*}[t]
\includegraphics[width=\textwidth, height=0.6\textwidth]{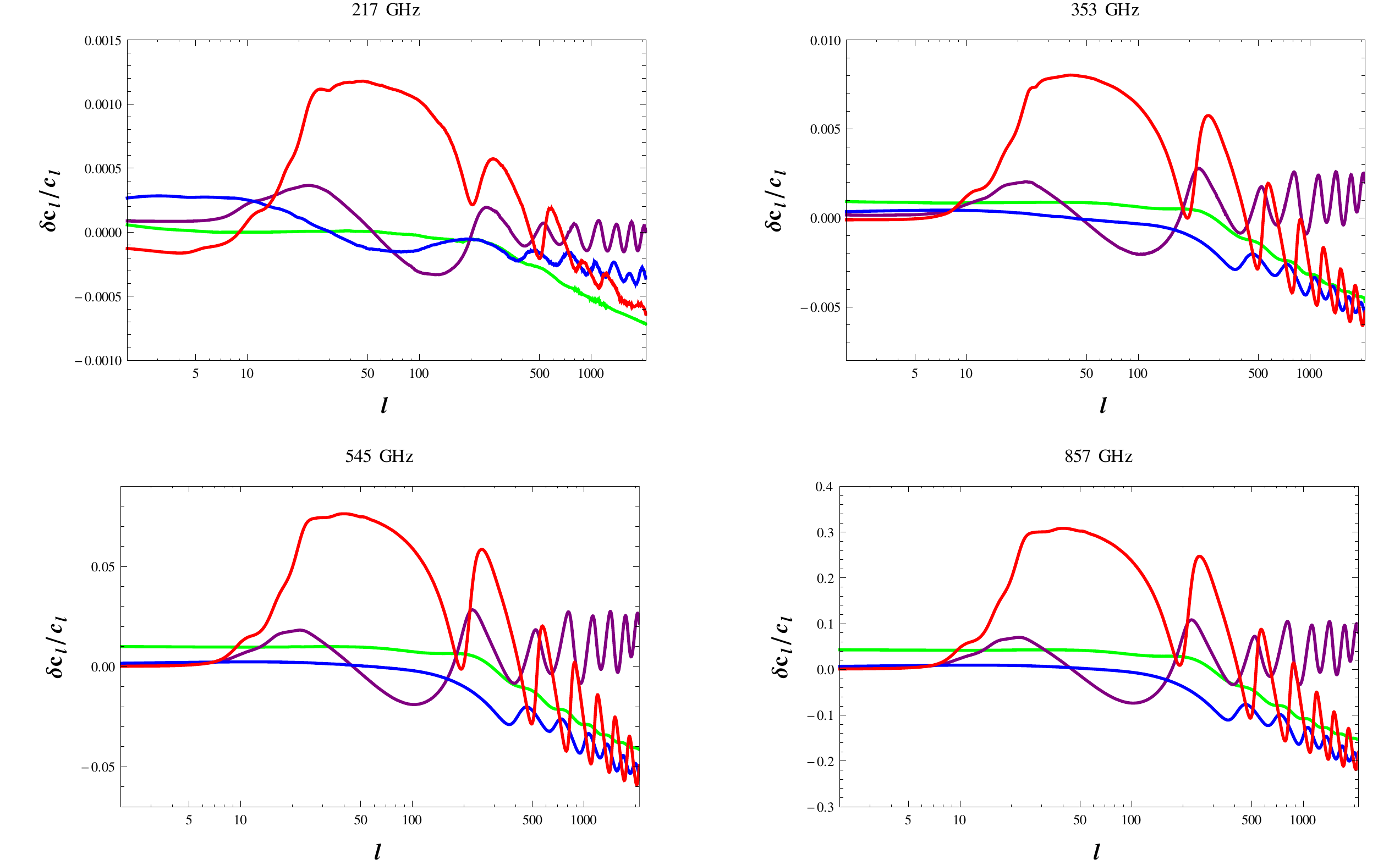}
\centering
\caption{Shown are a fractional measure, $\delta C_l^{XY}/\sqrt{C_l^{XX}C_l^{YY}}$, of the change $\delta C_l^{XY}$ in (lensed) scalar CMB anisotropy spectra due to Rayleigh scattering. The blue, red, green and purple are for the temperature, E-polarization, B-polarization from lensing and TE cross-correlation spectra respectively.}
\label{deltacl}
\end{figure*}

Using Eq. \ref{clXY}, the relative difference in the (lensed) scalar CMB power spectra due to Rayleigh scattering is calculated for four different frequencies and presented in Figure \ref{deltacl}. As expected, the relative difference in CMB power spectrum is bigger for higher frequencies. In the limit of very low frequencies the only modification in these power spectra arises from the increase in the total baryon-photon coupling due to Rayleigh scattering which is of order $0.05\%$.

On small scales, Rayleigh scattering leads to damping of both temperature and polarization anisotropies. Rayleigh scattering increases the rate of photon-baryon interaction and hence it reduces the photon-diffusion length. Since the amplitude of Silk damping depends on the integrated photon-diffusion length, it is also reduced by Rayleigh scattering. But there is another reason why the small-scale anisotropies are more damped in the presence of Rayleigh scattering. The damping factor at a given wave number is weighted by the photon visibility function. As we have seen above, adding Rayleigh scattering shifts the visibility function toward lower redshifts where Silk damping is more important and as a result, the anisotropy spectra at small scale decreases. 

We also find Rayleigh scattering leads to a boost in large-scale E-polarization. The reason for this is that the low-multipole polarization signal is sourced by the CMB quadrupole. Since the visibility function is shifted toward later time, where the quadrupole is larger, by Rayleigh scattering the low-multipole E-polarization signal is increased. In contrast, the effect of Rayleigh scattering on the lensing B modes is significantly smaller at low-multipole because these modes are produced by the gravitational lensing of E modes from a wide range of scales, so the Rayleigh contribution for them partly averages out.

Another effect worth noting is that, the oscillations of $\delta C_l/C_l$ show that the peaks in anisotropy spectra are shifted in the presence of Rayleigh scattering. Since the photon cross section is frequency dependent, the location of the surface of last scattering $\tau_*^{R+T}$ will depend on frequency too and the higher the frequency, the bigger $\tau_*^{R+T}(k,\nu)$. Therefore the sound horizon at the last scattering,
\be
 r_s^{R+T}=\int_0^{\tau_*^{R+T}} c_sd\tau,
\ee
will be larger than the sound horizon at last scattering when we only include the Thomson scattering $r_s^T$  and it will increase with increasing frequencies. The shift in the location of the peaks will be 
\be
\delta l/l=\delta k/k=1-r_s^{R+T}(\tau_*^{R+T})/r_s^T(\tau_*^T)
\ee
in the direction of decreasing $l$.
 
\section{Rayleigh Distorted Statistics}
\label{S_independent_variables}
Since the terms in the expansion of temperature and E-polarization perturbations, Eqs. \ref{Taylor1} and \ref{Taylor2}, fall off quickly like $(ah\nu/a_*{\rm Ry})^2$ only the two leading terms play an important role at frequencies smaller than 800 GHZ. We therefore effectively need two sets of random variables to describe the statistics of temperature and E-polarization. In this section we find a compressed representation of the power spectra for independent random variables. First we introduce the antenna temperature which is defined as
 \be
 T_{\rm ant}(\nu)=2\pi\nu f(\nu),
 \ee
 where $f(\nu)$ is the photon phase space distribution function and $\nu$ is the frequency. For the CMB, the antenna temperature has the form
 \be
 \frac{T_{\rm ant}(\nu)}{T}=\frac{h\nu/k_BT}{e^{h\nu/k_BT}-1}+\Theta \frac{(h\nu/k_BT)^2e^{h\nu/k_BT}}{(e^{h\nu/k_BT}-1)^2}.
 \ee
 The first term is the monopole which does not interest us here and we ignore it. The second term gives the spectral shape of CMB anisotropies. Keeping only the first two non-zero terms in Eqs. \ref{Taylor1} and \ref{Taylor2}, the antenna temperature for the CMB is
 \be
 \frac{T^X_{\rm ant}(\nu)}{T}=\Theta_X^{(0)} F^{(0)}(\nu)+\Theta_X^{(4)}F^{(4)}(\nu),
 \ee
 where $F^{(0)}(\nu)=\frac{(h\nu/k_BT)^2e^{h\nu/k_BT}}{(e^{h\nu/k_BT}-1)^2}$ is the black body shape function and $F^{(4)}(\nu)=(\frac{h\nu}{Ry})^4F^{(0)}(\nu)$ is the shape function for the Rayleigh signal and X is either I for intensity perturbations or E for E-polarization perturbations. The angular power spectrum covariance matrix for the antenna temperature is
 \bea
 C_l^{XX}(\nu,\nu')&&=C_l^{XX(00)}F^{(0)}(\nu)F^{(0)}(\nu')\nonumber\\
 +&&C_l^{XX(04)}(F^{(0)}(\nu)F^{(4)}(\nu')+F^{(4)}(\nu)F^{(0)}(\nu'))\nonumber\\
 +&&C_l^{XX(44)}F^{(4)}(\nu)F^{(4)}(\nu').
 \eea
 This structure indicates that $T_{\rm ant}(\nu)$ and $T_{\rm ant}(\nu')$ are correlated to each other but are not perfectly correlated like in the standard thermal case. We diagonalize the anisotropy spectrum in frequency space for a given $X \in \{I,E\}$ to obtain the two uncorrelated eigenvalues:
 \begin{widetext}
 \bea
 \lambda_{1,2}^{XX}(l)&=&[C_l^{XX(00)}G^{00}+2C_l^{XX(04)}G^{04}+C_l^{XX(44)}G^{44}\\
 &\pm & \sqrt{(C_l^{XX(00)}G^{00}+2C_l^{XX(04)}G^{04}+C_l^{XX(44)}G^{44})^2-4((C_l^{XX(04)})^2-C_l^{XX(00)}C_l^{XX(44)})((G^{04})^2-G^{00}G^{44})}]/2,\nonumber
 \eea 
 \end{widetext}
 where $G^{ij}= \int F^{(i)}(\nu)F^{(j)}(\nu)d\nu $. The two orthogonal eigenvectors are
 \bea
 v_{1,2l}^{X}(\nu)&=&N_{1,2}^X[(C_l^{XX(04)}\lambda_{1,2}^{XX}+C_l^{XX(00)}C_l^{XX(44)}G^{04}\\
 &-&(C_l^{XX(04)})^2G^{04})F^{(0)}(\nu)+(C_l^{XX(44)}\lambda_{1,2}^{XX}\nonumber\\
 &-&C_l^{XX(00)}C_l^{XX(44)}G^{00}+(C_l^{XX(04)})^2G^{00})F^{(4)}(\nu)],\nonumber
 \eea
 where $N_{1,2}^X$ are the normalization factor.
 If we expand the antenna temperature in terms of spherical harmonics,
 \be 
 T^X_{\rm ant}(\nu)/T=\sum _{l=1}^\infty \sum _{m=-l}^{l} a_{lm}^X Y_{lm},
 \ee
then we can write the coefficients $a_{lm}^X$ in the new basis spanned by the eigenvectors $\{v_{1l}^T(\nu),v_{2l}^T(\nu),v_{1l}^E(\nu),v_{2l}^E(\nu)\}$,
 \be
 a_{lm}^X=\alpha_{1lm}^Xv_{1l}^X(\nu)+\alpha_{2lm}^Xv_{2l}^X(\nu).
 \ee
 The covariance matrix in this new basis takes the compact form

 \be
 \textbf{C}_l\delta_{m,m'}=\left( \begin{array}{cc}
 \textbf{C}_l^I&\textbf{C}_l^{IE}\\
 \textbf{C}_l^{IE}&\textbf{C}_l^{E}
 \end{array}\right)\delta_{m,m'}=
 \ee
 \be
 \left( \begin{array}{cccc}
\langle\alpha_{1lm}^I\alpha_{1lm'}^I\rangle & 0 & \langle\alpha_{1lm}^I\alpha_{1lm'}^E\rangle & \langle\alpha_{1lm}^I\alpha_{2lm'}^E\rangle \\
0 & \langle\alpha_{2lm}^I\alpha_{2lm'}^I\rangle & \langle\alpha_{2lm}^I\alpha_{1lm'}^E\rangle & \langle\alpha_{2lm}^I\alpha_{2lm'}^E\rangle \\
\langle\alpha_{1lm}^E\alpha_{1lm'}^I\rangle & \langle\alpha_{1lm}^E\alpha_{2lm'}^I\rangle & \langle\alpha_{1lm}^E\alpha_{1lm'}^E\rangle & 0 \\
\langle\alpha_{2lm}^E\alpha_{1lm'}^I\rangle & \langle\alpha_{2lm}^E\alpha_{2lm'}^I\rangle & 0 & \langle\alpha_{2lm}^E\alpha_{2lm'}^E\rangle    \end{array} \right).\nonumber
 \ee
 
 \begin{figure*}[b]
\includegraphics[width=\textwidth, height=1.175\textwidth]{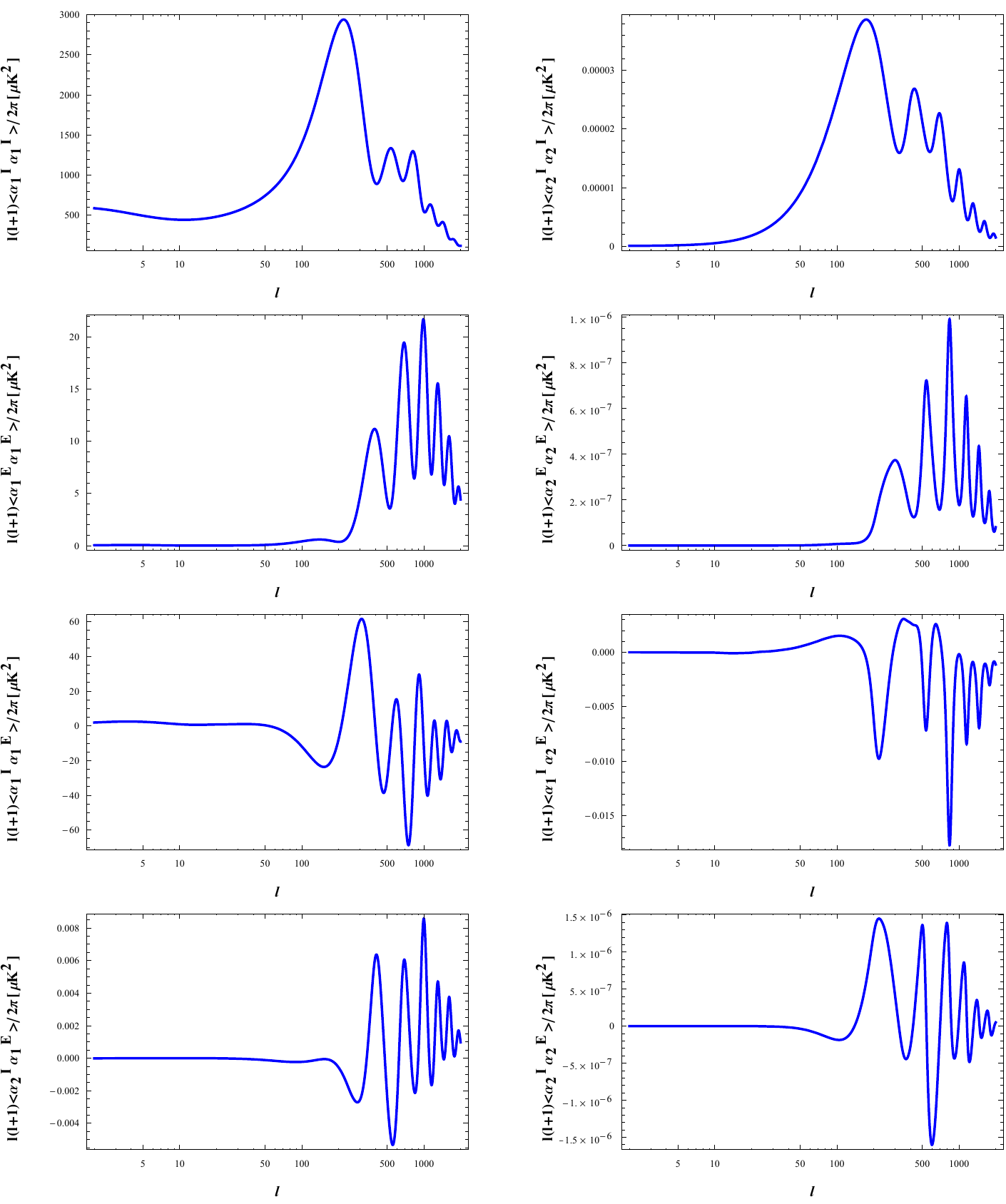}
\centering
\caption{The eight non-zero power spectra in the Rayleigh distorted CMB covariance matrix as a function of l. The first eigenvalues of intensity and polarization are almost proportional to the primary thermal signal and the second eigenvalues of intensity and E-polarization are purely Rayleigh signals which are uncorrelated to the first eigenvalues.}
 \label{elementsCl}
\end{figure*}
Using this diagonalization, we reduced the number of power spectra needed to describe the theoretical CMB covariance matrix from 10 to 8. These 8 non-zero elements in the covariance matrix are shown in Figure \ref{elementsCl}. $\langle\alpha_{1lm}^I\alpha_{1lm'}^I\rangle$ and $\langle\alpha_{1lm}^E\alpha_{1lm'}^E\rangle$ are almost proportional to the primary thermal signal (no Rayleigh scattering included) and we call them the primary temperature and polarization signal. The second eigenvalues of intensity and polarization spectra $\langle\alpha_{2lm}^I\alpha_{2lm'}^I\rangle$ and $\langle\alpha_{2lm}^E\alpha_{2lm'}^E\rangle$, which are due purely to Rayleigh scattering and uncorrelated to the first eigenvalues, we call the Rayleigh intensity and E-polarization signal. Note that since intensity and E-polarization perturbations must be separately diagonalized their eigenvectors are not orthogonal to each other. Thus all possible temperature-polarization cross-spectra are non-zero and present in Figure \ref{elementsCl}.
\section{Detectibility}
\label{S_detectibility}

Measurement of the Rayleigh signal is very challenging since at high frequencies that Rayleigh scattering becomes important, there are very few photons and very high levels of foreground contamination including Galactic dust, and the cosmic infrared background (CIB). Yet if many high frequency channels are measured in future CMB missions, in principle, foregrounds can be removed. The reason for this is that the spectral shape of foregrounds are different from one another and from the spectral shape of the Rayleigh signal. In addition, the Rayleigh power spectrum looks very different from all the foregrounds since it's oscillatory and it spans the full range of scales whereas most of the foregrounds are important either at lower or higher l values. For example, the CIB and thermal SZ have small amplitude at large scales but larger amplitude at smaller scales, while Galactic dust is important at lower l and is less so at higher l.  
 A future CMB mission that could be a candidate for detecting the Rayleigh signal is one similar to the proposed PRISM experiment \cite{PRISM}, which has many high frequency bands with more than 7000 detectors. In this section we take a PRISM-like experiment as an example of what capabilities a next generation CMB satellite might have and  explore the detectibility of the Rayleigh signal with this experiment.
\subsection{Signal to noise ratio of Rayleigh signal}

Our goal is to find the signal-to-noise ratio for the 8 non-zero elements of the CMB covariance matrix. As an example, we use the foreground removal method described in Ref.~\cite{Tegmark2000} and closely follow its notation. In this method, the foregrounds are treated as an additional source of noise which is correlated between frequency channels. If the frequency dependence, the scale dependence and also the variation in frequency dependence across sky are known for each physical component of foregrounds, this leads to a natural way of removing them. 

Let's say that our experiment has F frequency channels. The F-dimensional vectors $\textbf{a}_{lm}^I$ and $\textbf{a}_{lm}^E$, which are the measured multipoles at F different frequencies, are assumed to be composed of signal plus noise: 
 \be
 \textbf{y}_{lm}=\textbf{A}_l\textbf{x}_{lm}+\textbf{n}_{lm},
 \label{LinearEquation} 
 \ee
 \be
 \textbf{y}_{lm}=\left( \begin{array}{c} 
 \textbf{a}_{lm}^I\\
 \textbf{a}_{lm}^E
 \end{array} \right),\hspace{0.7cm} \textbf{x}_{lm}=\left( \begin{array}{c} 
 \alpha_{1lm}^I\\
 \alpha_{2lm}^I\\
 \alpha_{1lm}^E\\
 \alpha_{2lm}^E
 \end{array} \right),
 \ee
\be
\textbf{A}_l=\left( \begin{array}{cccc}
\textbf{v}_{1l}^I(\nu)&\textbf{v}_{2l}^I(\nu)&0&0\\
0&0&\textbf{v}_{1l}^E(\nu)&\textbf{v}_{2l}^E(\nu)
\end{array} \right).
\ee
 $\textbf{A}_l$ is the $2F\times  4$ scan strategy matrix for a given (l,m). $\textbf{n}_{lm}$ is the sum of detector noise and K different foregrounds components such as Galactic dust, synchroton emission or CIB. The covariance matrix for the noise is obtained by
 \be
 \textit{\textbf{N}}_l=\left( \begin{array}{cc}
 \textbf{N}_l^I&\textbf{N}_l^{IE}\\
 \textbf{N}_l^{IE}&\textbf{N}_l^E
 \end{array} \right),
 \ee  
 where $\textbf{N}_l^X=\sum_{k=1}^{K+1}\textbf{C}_l^X(k)$ is a $F\times F$ matrix. $\textbf{C}_l^X(k=1)$ is the covariance matrix for detector noise, and $\textbf{C}_l^X(k)$ is the angular power spectrum for different foreground components. 
 
To see how accurately we can remove the foregrounds and measure the CMB power spectra $\textbf{x}_{lm}$, we need to invert the noisy linear problem of Eq. \ref{LinearEquation}. It's shown in Ref.~\cite{Tegmark1997} that the minimum-variance estimate of the $\textbf{x}_{lm}$ is \mbox{$\tilde{\textbf{x}}_{lm}=\textit{\textbf{W}}_l^t\textbf{y}_{lm}$} where
 \bea
 \textit{\textbf{W}}_l&=&\textit{\textbf{N}}_l^{-1}\textbf{A}_l[\textbf{A}_l^t\textit{\textbf{N}}_l^{-1}\textbf{A}_l]^{-1}\nonumber\\
 &=&\left( \begin{array}{cccc}
 \textbf{w}_{1l}^I&\textbf{w}_{2l}^{I}&\textbf{w}_{1l}^E&\textbf{w}_{2l}^{E}\\
 \textbf{w}_{1l}^{I'}&\textbf{w}_{2l}^{I'}&\textbf{w}_{1l}^{E'}&\textbf{w}_{2l}^{E'}
 \end{array}\right).
 \eea
 $\textbf{w}_{il}^X$ are the F-dimensional weight vectors where
\bea
\tilde{\alpha}_{ilm}^I&=&\textbf{w}_{il}^{It}\textbf{a}_{lm}^I+\textbf{w}_{il}^{I't}\textbf{a}_{lm}^E,\nonumber\\
\tilde{\alpha}_{ilm}^E&=&\textbf{w}_{il}^{Et}\textbf{a}_{lm}^E+\textbf{w}_{il}^{E't}\textbf{a}_{lm}^I.
\eea 
 The weight vectors are different for each l-value, so that at each angular scale, the frequency channels with smaller foregrounds contribution have more weight. 
 
 The estimated solution $\tilde{\textbf{x}}_{lm}$ is unbiased such that $\langle\tilde{\textbf{x}}_{lm}\rangle=\textbf{x}_{lm}$ and the covariance matrix of the pixel noise $\boldsymbol\varepsilon_{lm}=\tilde{\textbf{x}}_{lm}-\textbf{x}_{lm}$ is $\boldsymbol\Sigma_{l}\delta_{m,m'} =\langle\boldsymbol\varepsilon_{lm}\boldsymbol\varepsilon_{lm'}^t\rangle$ where
 \be
\boldsymbol\Sigma_{l} =[\textbf{A}_l^t\textit{\textbf{N}}_l^{-1}\textbf{A}_l]^{-1}=\left(\begin{array}{cc}
\tilde{\textbf{N}}_l^I&\tilde{\textbf{N}}_l^{IE}\\
\tilde{\textbf{N}}_l^{IE}&\tilde{\textbf{N}}_l^E
\end{array}\right).
 \ee
 Here $\tilde{\textbf{N}}_l^I$, $\tilde{\textbf{N}}_l^E$  and $\tilde{\textbf{N}}_l^{IE}$ are $2\times 2$ cleaned power spectrum matrices of the non-cosmic signals. The covariance matrix of our estimate $\tilde{\textbf{x}}_{lm}$ is
\be
\tilde{\textbf{C}}_l\delta_{m,m'}=\langle\tilde{\textbf{x}}_{lm}^\ast\tilde{\textbf{x}}_{lm}^t\rangle=\left( \begin{array}{cc}
\tilde{\textbf{C}}_l^I&\tilde{\textbf{C}}_l^{IE}\\
\tilde{\textbf{C}}_l^{IE}&\tilde{\textbf{C}}_l^E
\end{array}\right)\delta_{m,m'},
\ee
where $\tilde{\textbf{C}}_l^X=\textbf{C}_l^X+\tilde{\textbf{N}}_l^X$ is the total power spectrum in the cleaned maps. To find how accurately we can measure any of the eight non-zero element of cosmic power spectrum, we must compute the $8\times 8$ Fisher matrix:
\be
\textbf{F}_{l\alpha \beta}=\frac{1}{2}Tr[\tilde{\textbf{C}}_l^{-1}\frac{\partial \tilde{
\textbf{C}}_l}{\partial \alpha}\tilde{\textbf{C}}_l^{-1}\frac{\partial \tilde{
\textbf{C}}_l}{\partial \beta}],
 \label{Fisher}
\ee 
where $\alpha$ and $\beta$ could be any of the 8 non-zero elements. Up this point, we have used only one multipole $\textbf{x}_{lm}$ to calculate the Fisher matrix, but for each l-value we have $(2l+1)f_{\rm sky}$ independent modes where $f_{\rm sky}$ is the fraction of sky covered. Therefore the full Fisher matrix is $(2l+1)f_{\rm sky}$ times what we calculated in Eq \ref{Fisher}. Inverting this matrix gives the constraints on the 8 non-zero elements of the cosmic covariance matrix.

\begin{figure*}[t]
\includegraphics[width=\textwidth, height=1.175\textwidth]{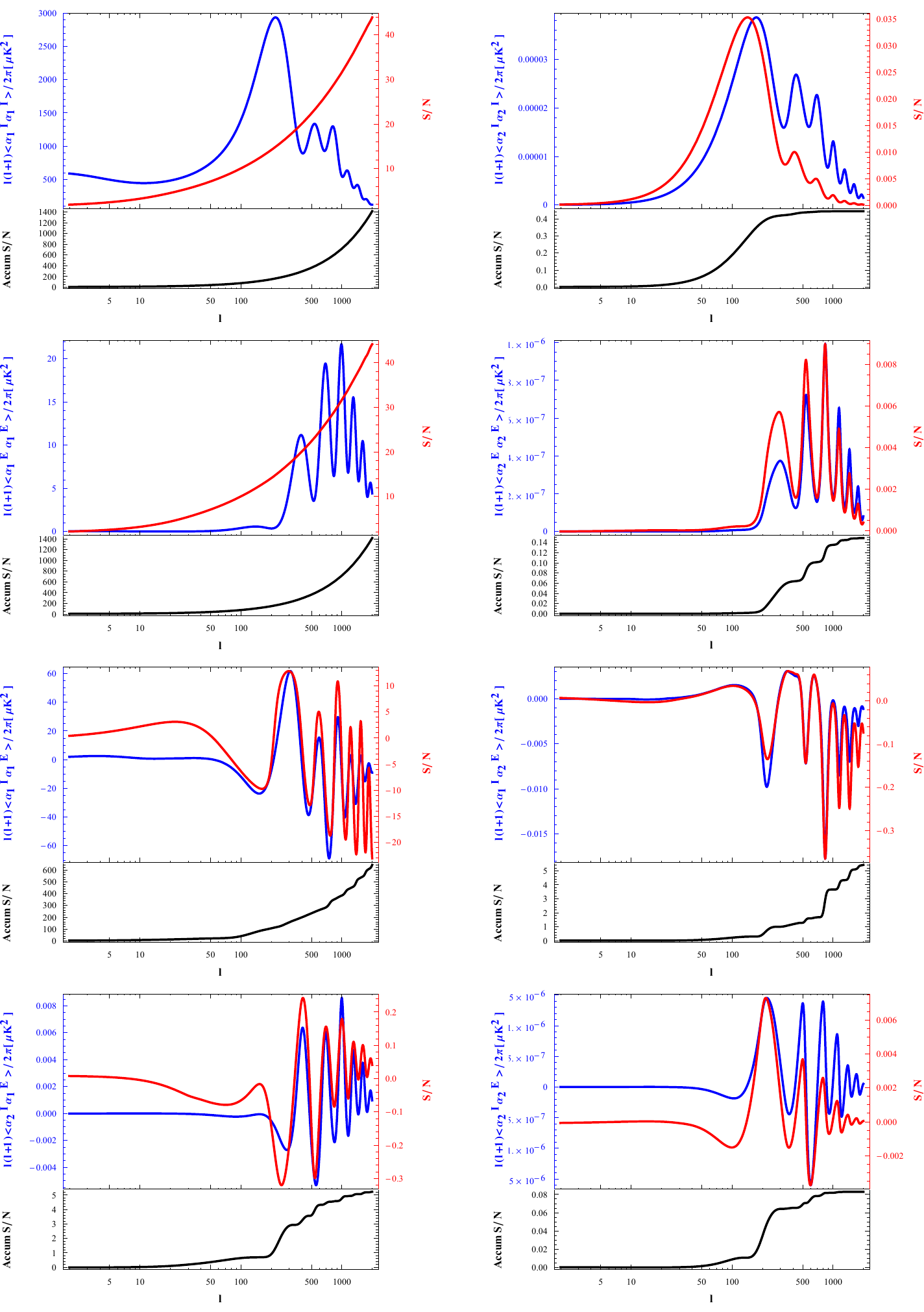}
\centering
\caption{The eight non-zero elements of the Rayleigh-distorted CMB covariance matrix and their signal-to-noise ratio at each l as well as the accumulative signal-to-noise ratio for the PRISM-like experiment.}
 \label{SN}
\end{figure*}

We compute this Fisher matrix for a PRISM-like experiment with the same frequency channels between 30GHz and 800GHz as PRISM. For the noise, we choose the resolution to be 1 arc min and the sensitivity to be 1nK for channels with frequencies less than 500GHz and 10nK for channels with frequencies higher than 500GHz. For the dominant foregrounds components, the temperature and E-polarization power spectra of Galactic dust and the temperature power spectra of CIB, we used the power spectra given in a series of Planck papers \cite{PlanckCMB, PlanckDust,PlanckDustPolarization, PlanckCIB}. For other foregrounds components which are subdominant for detecting the Rayleigh signal, we used the power spectra given in Table 2 of Ref.~\cite{Tegmark2000}. The eight non-zero elements and their signal-to-noise ratio for each l value as well as accumulative signal-to-noise ratio are plotted in Fig \ref{SN}.

Since the power spectra $\langle\alpha_{1lm}^I\alpha_{1lm'}^I\rangle$, $\langle\alpha_{1lm}^E\alpha_{1lm'}^E\rangle$ and $\langle\alpha_{1lm}^I\alpha_{1lm'}^E\rangle$ are almost the same as the primary thermal signal, their signal-to-noise ratio is huge. For the auto correlation of the primary temperature and E-polarization, the signal-to-noise ratio is almost equal to the cosmic variance limit up to $l=2000$. Among the remaining elements, $\langle\alpha_{1lm}^I\alpha_{2lm'}^E\rangle$ and $\langle\alpha_{2lm}^I\alpha_{1lm'}^E\rangle$ have larger accumulative signal-to-noise ratios and these two are detectable for this PRISM-like experiment. The accumulative signal-to-noise for $\langle\alpha_{1lm}^I\alpha_{2lm'}^E\rangle$ is almost 5.4 and for $\langle\alpha_{2lm}^I\alpha_{1lm'}^E\rangle$ is around 5.2. A detection of these Rayleigh-distorted statistics would be an interesting and non-trivial cross check of the CMB physics and the assumed cosmological model.

\subsection{Constraints on Cosmological Parameters}

There is independent information contained in the Rayleigh signal which might help to better constrain the cosmological parameters. To show how much potential information we can get from the Rayleigh signal, we consider an ideal experiment with no foregrounds and negligible detector noise so that the signal-to-noise ratios for both the primary and Rayleigh signals are cosmic-variance limited. To find the constraints on seven cosmological parameters, $\Omega_b, \Omega_c,\tau, n_s, A_s, H, Y_p$, we calculate the Fisher matrix using the standard equation:
\be
\textbf{F}_{ij}=\sum_l^{l_{\rm max}} (2l+1)f_{\rm sky}\frac{1}{2}Tr[\tilde{\textbf{C}}_l^{-1}\frac{\partial \tilde{
\textbf{C}}_l}{\partial p_i}\tilde{\textbf{C}}_l^{-1}\frac{\partial \tilde{
\textbf{C}}_l}{\partial p_j}],
\ee
where $p_i$ and $p_j$ could be any of the seven cosmological parameters considered. The constraints on cosmological parameters for the cosmic-variance limited experiment are presented in Table \ref{constraints}. Note that in this calculation we only included moments up to $l_{\rm max}=2000$. In principle the extra information contained in the Rayleigh sky is quite powerful. For instance, adding the Rayleigh signal potentially could help to improve the constraint on the helium fraction $Y_p$ by a factor of four. Furthermore, the fundamental limit on $n_s$ from the CMB only is less than $10^{-3}$ which could be of interest for inflation studies. 

\begin{table}
\begin{tabular}{|c|c|c|c|c|c|c|}
\hline
 & parameter values & Primary & Primary+Rayleigh  \\
 & Planck+WP & CV Limited & CV Limited  \\
\hline
 $\Omega_b$ & 0.02205 &  0.25657 & 0.10136  \\
 $\Omega_c$ & 0.1199 & 0.3570 & 0.1149 \\
 $\tau$ & 0.089 &  2.4033 & 1.0887 \\
 $n_s$ & 0.9603 & 0.2623 & 0.0950 \\
 $A_s$ & $2.1955\times 10^{-9}$ & 0.4009 & 0.1829 \\
 $H$ & 67.3 & 0.2667 & 0.0870 \\
 $Y_p$ & 0.24770 & 1.4288 & 0.3375 \\
\hline
\end{tabular}
\caption{The percentage constraints on cosmological parameters ($100\sigma_{p_i}/p_i$) for the ideal cosmic-variance limited case with and without accounting for the Rayleigh signal.}
\label{constraints}
\end{table} 
\begin{figure*}[b]
\includegraphics[width=\textwidth, height=1.15\textwidth]{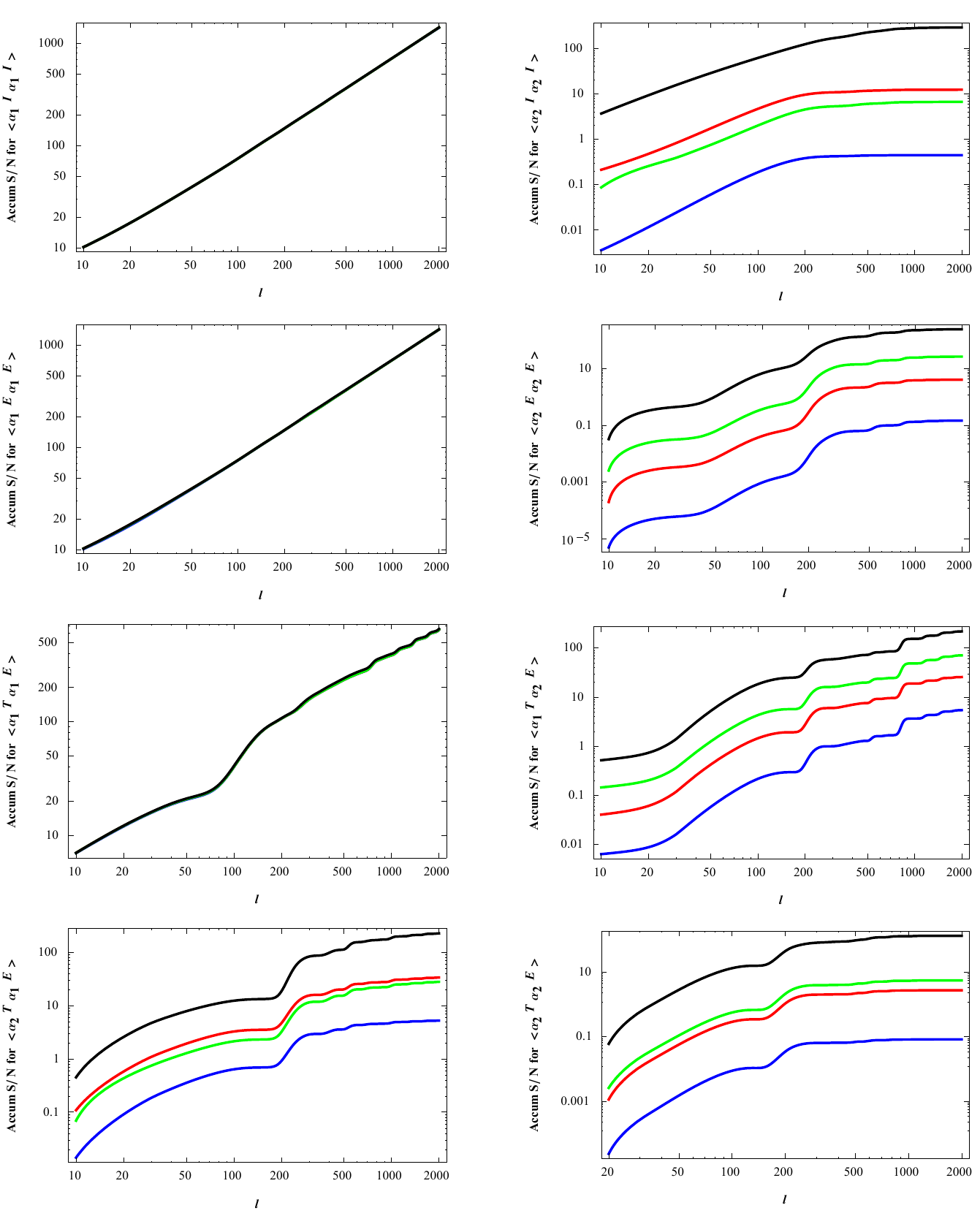}
\centering
\caption{ Accumulative signal-to-noise ratios for the eight non-zero elements of the CMB covariance matrix. The blue, red, green and black lines are the signal-to-noise ratios respectively for a PRISM-like experiment, for Case I: improved foregrounds removal method, for Case II : improved detector noise, and for Case III which combines Case I and II. }
 \label{AccumSN}
\end{figure*}

We also calculate how much of a constraint one can except for the PRISM-like experiment. In this case, although the Rayleigh signal is detectable, the Rayleigh signal adds very little constraining power for cosmological parameters as its accumulative signal-to-noise ratio is small. 

 It's also reasonable to ask how biased each cosmological parameter will be  by ignoring the Rayleigh scattering. These biases will move the central measured values of each parameter relative to  their actual values. The observed power spectrum is a sum of the primary power spectrum, Rayleigh power spectrum and generalized noise (including foregrounds)
 \be
\tilde{\textbf{C}}_l=\textbf{C}_l^{\rm Primary}+\textbf{C}_l^{\rm Rayleigh}+\tilde{\textbf{N}}_l.
\ee
 
 To calculate the bias, we need to find the difference between expectation value of the parameter estimator, $\langle\hat{p}_i\rangle$, and the true value $\overline{p}_i$,  using
 \be
\textbf{b}_i= \langle\hat{p}_i\rangle-\overline{p}_i = \textbf{F}^{(00)-1}_{ij}\textbf{B}_{j},
\ee
where $\textbf{F}^{(00)-1}_{ij}$ and $\textbf{B}_{j}$ are Fisher matrix and bias vector respectively for the power spectrum $\textbf{C}_l^{P}=\textbf{C}_l^{\rm Primary}+\tilde{\textbf{N}}_l$ 
\be
\textbf{F}^{(00)-1}_{ij}=\sum_l ^{l_{max}} (2l+1)\frac{1}{2}Tr[\textbf{C}_l^{P -1}\frac{\partial \textbf{C}_l^P}{\partial p_i}\textbf{C}_l^{P -1}\frac{\partial \textbf{C}_l^P}{\partial p_j}],
\ee
\be
\textbf{B}_j = \sum_l ^{l_{max}}(2l+1)\frac{1}{2}Tr[\textbf{C}_l^{P -1}\frac{\partial \textbf{C}_l^P}{\partial p_j}\textbf{C}_l^{P -1}\textbf{C}_l^{\rm Rayleigh}].
\ee

The biases (relative to standard deviation) introduced by ignoring the Rayleigh scattering for the PRISM-like experiment are $b_i/\sigma_i=\{-0.13,0.08,-0.06,-0.20,-0.02,-0.18,-0.28\}$ for the set of parameters $\{\Omega_b, \Omega_c,\tau, n_s, A_s, H, Y_p\}$. While these potential biases are worrisome and Rayleigh scattering should be incorporated into future analysis, they are still smaller than the forecast constraints on each parameter.

The potential constraints that could be achieved using a cosmic-variance limited experiment, motivate us to consider how larger signal-to-noise measurements might be made. 
\subsection{Improvements to signal to noise ratio}

There are a few ways to improve the signal-to-noise ratio of the Rayleigh signal and bring it closer to the idealized cosmic-variance limit. One is to have a more effective foreground removal method. The scheme we discussed assumes an isotropic power spectrum for each foreground component and aims to detect the signal in the presence of foregrounds using only this knowledge. Since Rayleigh scattering is more important at frequencies higher than 300GHz and at high frequencies the dominant foregrounds are Galactic dust and CIB, one might do a better job at foreground removal by measuring Galactic dust and CIB maps at very high frequency, (for example higher than 600GHz), and then extrapolating their spectrum and removing them at the map level from lower frequencies such as 300GHz or 400 GHz. While we will still be left with some residual foreground power spectra they should have a smaller amplitude than the original foreground power spectra. Furthermore, as long as the Rayleigh signal in not limited by cosmic variance, instead of probing the whole sky one could concentrate observing time on regions of the sky where foreground contamination is less. 

 Another way to enhance the signal-to-noise ratio is to  improve the experiment. To do so, we can either reduce the detector noise by having more detectors (better sensitivity) or by including more frequency channels so that we can model foregrounds with higher fidelity and remove them more effectively.

To examine how sensitive the signal-to-noise ratio of the Rayleigh signal is to each of these improvements, we study three cases: $\textbf{ Case I}.$  In the first case, we keep the specification of the experiment the same as our PRISM-like experiment but imagine a more effective foreground removal method. More specifically, in this case, by measuring the foregrounds at very high frequencies or optimizing observation to low foreground region, we assume we can remove most of the foregrounds spectra from lower frequencies and are left with only $5\%$ of the original foreground spectra as residuals. $\textbf{ Case II}.$  In the second case we use the same normal foreground levels but improve the specification of the experiment. For illustrating purposes we consider an extremely ambitious experiment with 50 frequency channels between 30 GHz and 800 GHz and a noise in each frequency channel of 0.01 nK. $\textbf{ Case III}.$  The third case is the combination of I and II. 
\begin{figure}[t]
\includegraphics[width=0.45\textwidth, height=0.7\textwidth]{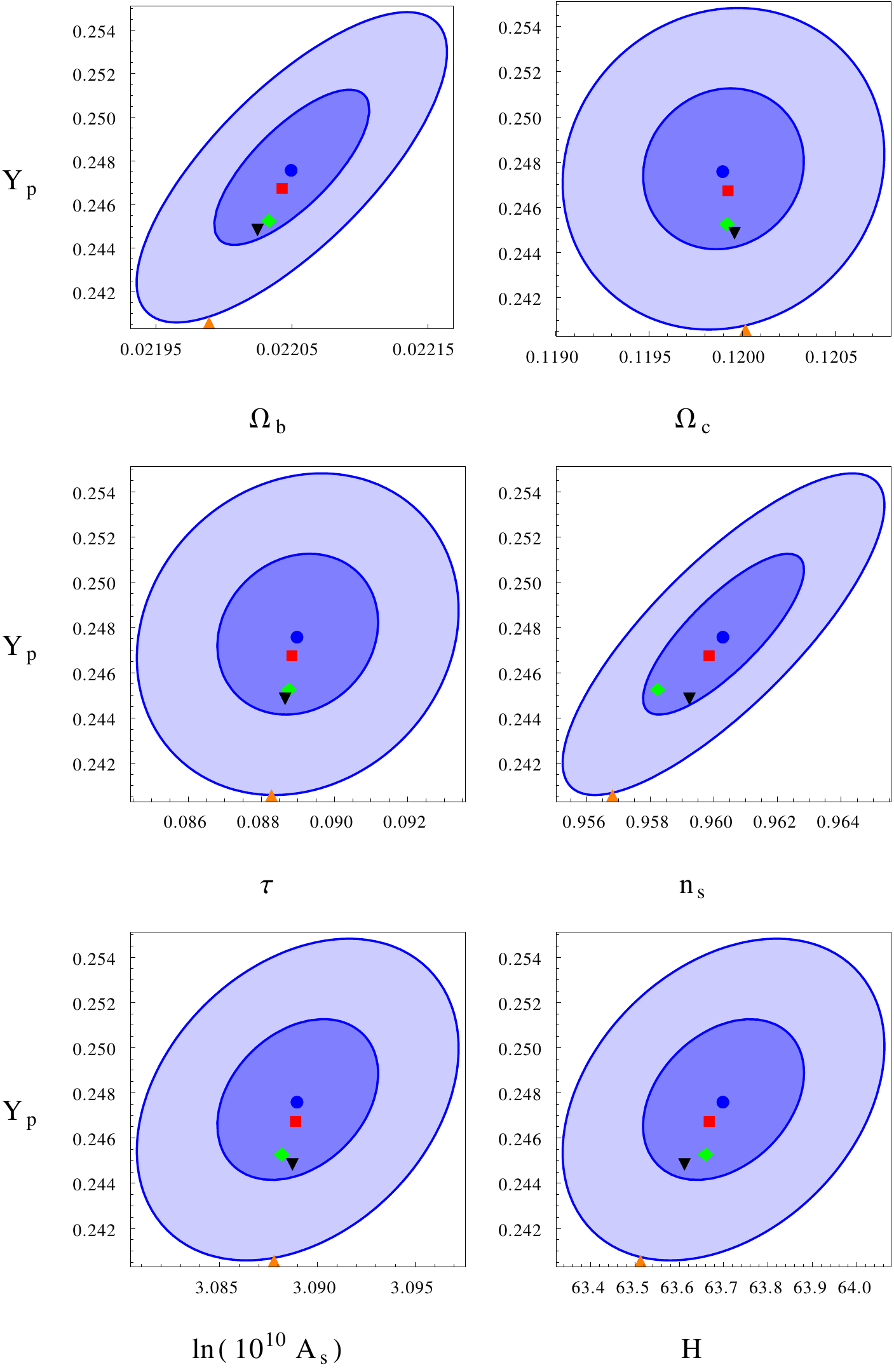}
\centering
\caption{The biases and constraints  on cosmological parameters that could potentially occur if one ignores the Rayleigh signal. The blue contours are the one-sigma and two-sigma constraints on parameters using only the primary signal centred at the fiducial value of the parameters. The red, green, orange and black dots represent the bias introduced by ignoring the Rayleigh signal respectively in PRISM-like experiment, Case I (improving foreground removal), Case II (reducing detector noise) and Case III (combination of both). }
 \label{biasPrimary}
\end{figure}

In Figure \ref{AccumSN}, we show the accumulative signal-to-noise ratios for the all eight non-zero elements of CMB covariance matrix for these improved cases. The blue, red, green and black lines are the signal-to-noise ratios respectively for a PRISM-like experiment, Case I, Case II and Case III. For example, the accumulative signal-to-noise ratio for the cross spectra between the primary temperature signal and Rayleigh E-polarization signal which was around 5 for the PRISM-like experiment, is amplified to 26 by improving the foregrounds removal method (Case I), to 71 by decreasing the detector noise (Case II) and to 218 by combining Case I and II (Case III). As can be seen from this graph, in Case III the accumulative signal-to-noise ratio of all the power spectra are greater than 100 and could provide us with valuable information about cosmological parameters.

\begin{figure*}[b]
\includegraphics[width=1\textwidth, height=1\textwidth]{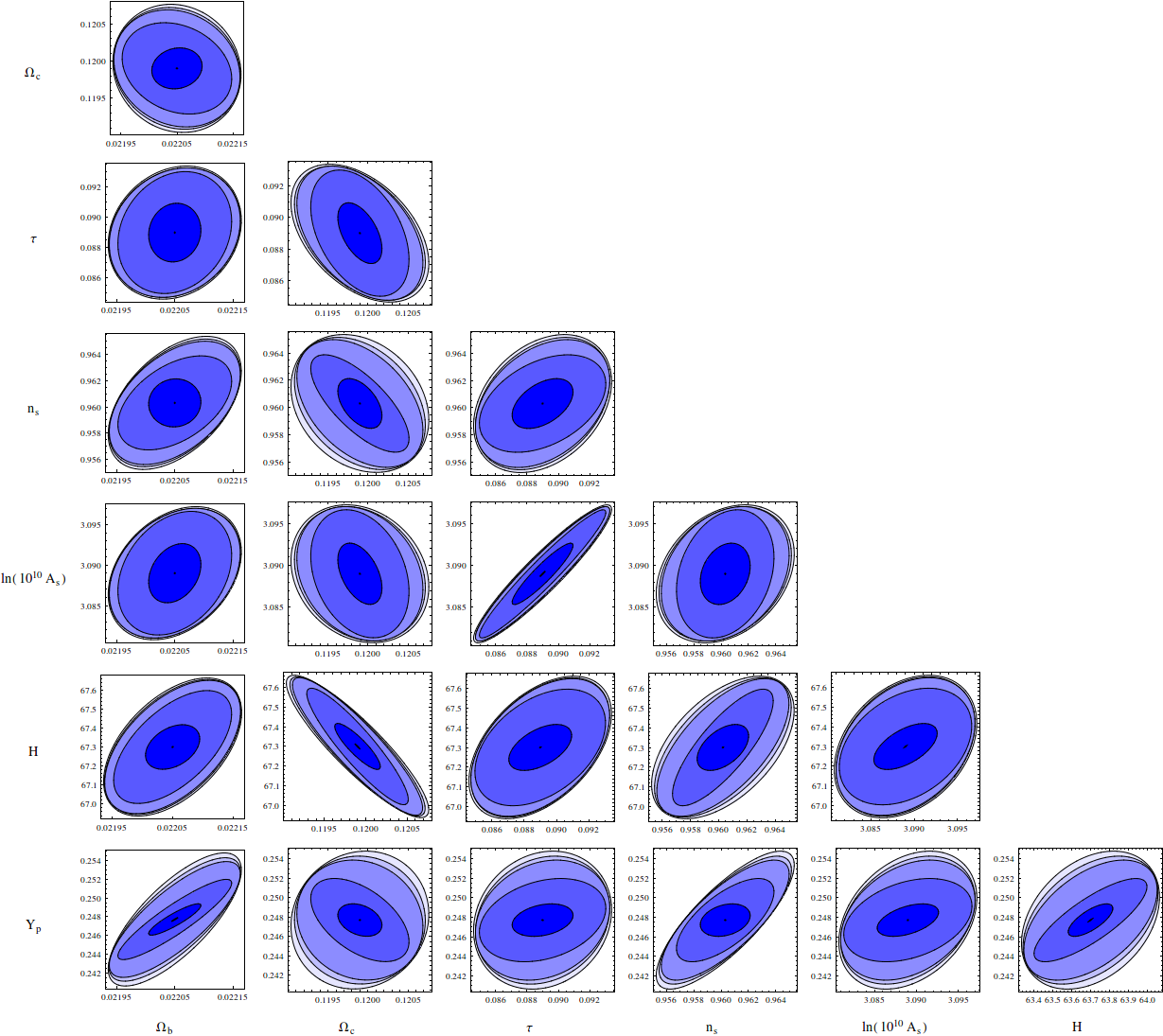}
\centering
\caption{The two-sigma constraints on cosmological parameters by considering both the primary and Rayleigh signal. The smallest and darkest contour represents the cosmic-variance limited case. The lighter contours show the Case III, Case II, Case I and the PRISM-like experiment respectively as we go from smallest-darkest to largest-lightest contours. Note that  the largest contours essentially delineate the conventional (primary only) cosmic variance limit, and smaller contours represent an improvement in parameter constraints beyond this limit. }
 \label{2sigma4cases}
\end{figure*}

The effects of improving the signal-to-noise ratio on cosmological parameters are illustrated in Figures \ref{biasPrimary} and \ref{2sigma4cases}. In Figure \ref{biasPrimary} we plotted the one-sigma and two-sigma constraints on cosmological parameters using only the primary signal. Since the signal-to-noise ratio for the primary signal is cosmic-variance limited in all the cases considered here, the constraints on the parameters remain the same for all cases. We also show the bias introduced by ignoring the Rayleigh signal in this Figure. In almost all the cases (save for one) the bias for each parameter is less than one sigma and only when the foreground contamination is large and the detector noise is small, Case II, we are left with biases larger than two sigma for some parameters.

In Figure \ref{2sigma4cases}, we plotted the two-sigma constraints on cosmological parameters using both the primary and Rayleigh signal and show that by improving the PRISM-like experiment, as we go through Case I, II and III, the constraints on parameters become smaller since the signal-to-noise ratio of the Rayleigh signal becomes larger. For instance, the percentage error on $Y_p$ in case III is half the constraint of the PRISM-like experiment.

\section{Conclusions}
\label{S_conclusion}

In this paper, we have calculated the effect of Rayleigh scattering on CMB temperature and polarization anisotropies as well as the impact on cosmic structure. We also have investigated the possibility of detecting the Rayleigh signal in the CMB. A new method was introduced to account for the frequency-dependence of the Rayleigh cross section by solving for a hierarchy of spectral distortion perturbations, which allows for an accurate treatment of Rayleigh scattering including its back-reaction on baryon perturbations with only a few spectral-distortion hierarchies.
We have found that Rayleigh scattering modifies the distribution of matter in the Universe at the $0.3\%$ level.
 
Since the Rayleigh cross section is frequency-dependent, the CMB temperature and polarization anisotropies depend on frequency too. For each frequency of interest, Rayleigh scattering reduces the $C_l$ power spectrum at high l multipoles because the visibility function shifts to lower redshifts when the Silk damping is more important. For reference, at 857 GHz, the highest frequency of the Planck experiment, both temperature and E-polarization anisotropies decrease as much as $20\%$ near $l\sim 1000$ and at 353 GHz they decrease as much as $0.6\%$. Low-multipole E-polarization anisotropies increase because the visibility function shifts toward later time when the CMB quadrupole is larger. The increase in E-polarization signal at $l\sim 50$ is $35\%$ at 857 GHz and $0.8\%$ at 353 GHz.

We showed that due to these distortions, the primary intensity and E-polarization power spectra at different frequencies are not perfectly correlated with each other like in standard treatments of the CMB. Furthermore we have found, to a very good approximation, we need two sets of random variables to completely describe the statistics of primordial intensity and E-polarization patterns on the sky we observe. There is a second Rayleigh-distorted CMB sky beyond the primary CMB sky that contains additional information. We have determined a compressed representation of the joint power spectra of these two temperature/intensity and E-polarization skies.

Detecting the Rayleigh signal is very challenging because at high frequencies the number of CMB photons is low and the signal is contaminated by foregrounds. However since both the spectral shape and power spectra of the Rayleigh sky are different from all the foregrounds, the Rayleigh signal might be detectable if many high frequency channels are included in future CMB missions. We have shown that with a PRISM-like experiment that has many frequency bands, and using a simple power spectrum based foregrounds removal method, the cross spectrum between the primary E-polarization and Rayleigh temperature signal and the cross spectrum between the primary temperature and Rayleigh E-polarization signal should be detectable with accumulative signal-to-noise ratio of 5.2 and 5.4 respectively. 

Measuring the Rayleigh signal could provide powerful constraints on cosmological parameters including the helium fraction and scalar spectral index. A more ambitious experiment either observing in low foreground contaminated regions or using a more sophisticated foreground removal method might detect the Rayleigh CMB sky at high signal-to-noise. This would tighten CMB constraints on cosmological parameters beyond what was, even in principle, previously thought possible.

\acknowledgements 

We thank Graeme Addison, Mark Halpern, Gary Hinshaw, Antony Lewis and Douglas Scott for discussion. 
The
research of EA is supported in part by a Four Year Doctoral Fellowship from the University of British Columbia and  the Natural Sciences and Engineering Research Council (NSERC) of Canada.  KS is supported in part by a NSERC of Canada Discovery Grant.  C.H. is supported by the United States Department of Energy under contract DE-FG03-02-ER40701, the David and Lucile Packard Foundation, the Alfred P. Sloan Foundation, and the Simons Foundation.

\noindent {\bf Note:} After implementing this method for including Rayleigh scattering in cosmological perturbation calculations, Ref.~\cite{Lewis} appeared discussing an alternative method. 
We have verified these distinct methods agree very well, with remaining differences consistent with the size of the baryon back-reaction effects we find here Ref.~\cite{Lewis2}.

\end{document}